\documentclass[journal]{IEEEtran}
\ifCLASSINFOpdf
\else
\fi
\usepackage{amsfonts}
\usepackage{amsmath,amsfonts,amssymb,epsfig,subfigure,cite,url,multirow,booktabs}
\usepackage{afterpage,mathrsfs,algorithm,algorithmic}
\usepackage{amsthm}

\newtheorem{definition}{\textbf{Definition}}
\newtheorem{lemma}{\textbf{Lemma}}

\newtheorem{proposition}{\textbf{Proposition}}
\newtheorem{remark}{\textbf{Remark}}

\begin{document}
%
\title{Integrated Communications and Security: RIS-Assisted Simultaneous Transmission and Generation of Secret Keys }
%

%

\author{Ning Gao,~\IEEEmembership{Member,~IEEE,}~Yuze Yao,~\IEEEmembership{Student Member,~IEEE,}~Shi Jin,~\IEEEmembership{Fellow,~IEEE,}\\~Cen Li, and~Michail Matthaiou,~\IEEEmembership{Fellow,~IEEE}

\thanks{This work was supported in part by the National Science
Foundation of China (NSFC) under Grants 62371131, in part by the National Key Research and Development Program of China under Grant 2024YFE0200703, in part by the Guangdong Provincial Key
Laboratory of Novel Security Intelligence Technologies under
Grant 2022B1212010005, in part
by the Start-up Research Fund of Southeast University under Grant 4009012307, and in part by the program of Zhishan Young Scholar of Southeast University under Grant 2242024RCB0030. The work of M. Matthaiou was supported in part by the European Research Council (ERC) under the European Union's
Horizon 2020 research and innovation programme (grant agreement No. 101001331) and in part by the U.K. Engineering and Physical Sciences Research Council (EPSRC) (grant No. EP/X04047X/1).}
\thanks{N. Gao is with the School of Cyber Science
and Engineering, Southeast University, Nanjing 210096, China, and also with the Guangdong Provincial Key
Laboratory of Novel Security Intelligence Technologies, Shenzhen 518055, Guangdong, China (e-mail:
ninggao@seu.edu.cn).}

\thanks{Y. Yao and C. Li are with the School of Cyber Science
and Engineering, Southeast University, Nanjing 210096, China (e-mail: yuzeyao@seu.edu.cn; Licen@seu.edu.cn).}
\thanks{S. Jin is with the National Mobile Communications
Research Laboratory, Southeast University, Nanjing 210096, China (e-mail: jinshi@seu.edu.cn).}
\thanks{
M. Matthaiou is with the Centre for Wireless Innovation (CWI), Queen’s University Belfast, Belfast BT3 9DT, U.K. (e-mail: m.matthaiou@qub.ac.uk).}
}
\markboth{}%
{Shell \MakeLowercase{\textit{et al.}}: Bare Demo of IEEEtran.cls for IEEE Journals}
%



\maketitle

\begin{abstract}
We develop a new integrated communications and
security (ICAS) design paradigm by leveraging the concept of reconfigurable intelligent surfaces (RISs). In particular, we propose RIS-assisted simultaneous transmission and secret key generation by sharing the RIS for these two tasks. Specifically, the legitimate transceivers intend to jointly optimize the data transmission rate and the key generation rate by configuring the phase-shift of the RIS in the presence of a smart attacker. We first derive the key generation rate of the RIS-assisted physical layer key generation (PLKG). Then, to obtain the optimal RIS configuration, we formulate the problem as a secure transmission (ST) game and prove the existence of the Nash equilibrium (NE), and then derive the NE point of the static game. For the dynamic ST game, we model the problem as a finite Markov decision process
and propose a model-free reinforcement learning approach to obtain the NE point. Particularly, considering that the legitimate transceivers cannot obtain the channel state information (CSI) of the attacker in real-world conditions, we develop a deep recurrent Q-network (DRQN) based dynamic ST strategy to learn the optimal RIS configuration. The details of the algorithm are provided, and then, the system complexity is analyzed. Our simulation results show that the proposed DRQN based dynamic ST strategy has a better performance than the benchmarks even with a partial observation information, and achieves ``one time pad" communication by allocating a suitable weight factor for data transmission and PLKG.
\end{abstract}


%
\IEEEpeerreviewmaketitle
\section{Introduction}
\IEEEPARstart{W}{ith} the evolution of the sixth-generation (6G) wireless networks, we are moving towards the Internet of Everything (IoE) era. The frequent access of the mobile edge devices, the tremendous connectivity of the heterogeneous Internet of Things (IoT) devices and the peak throughput requirements of Tbps level are emphatically underscoring the unprecedented importance of data security \cite{9482503}. Traditionally, the data integrity, confidentiality and non-repudiation rely mainly on the key encryption form the upper layers. The classical key generation techniques, such as Diffie-Hellman, depend typically on the complexity of the mathematical problem. However, the large number of IoT devices escalates the difficulty of key distribution and update via the public key infrastructures (PKI). Furthermore, with the discovery of Shor's algorithm, the traditional cryptographic protocols face the risk of being cracked by quantum computers.

In recent years, physical layer key generation (PLKG) is regarded as a very promising technology to address the key distribution and update by using the endogenous stochastic
and reciprocal properties of the wireless channel \cite{8883129}. Due to the fact that the key source comes from the physical propagation channel, PLKG stands out as the most promising strategy for achieving Shannon’s perfect ``one time pad" communication. The standard process of PLKG can be divided into four steps, which are: channel sounding, quantization, information reconciliation and privacy amplification \cite{8883129,7539590,7448884}. First, in a channel coherence time, the transceivers probe the channel and extract the channel reciprocity feature by sending the pilot sequences. Next, the channel features are independently quantified into binary bit sequences at transceivers, which are used as raw bit sequences for key generation. Note that due to quantization
accuracy, noise and imperfect synchronization, etc., the mismatched bits in the raw bit sequences should be corrected by
using an error correcting code, i.e., low density parity
check (LDPC) code, during information reconciliation. Finally, privacy amplification is necessary to remove the possible information leakage of the raw bit sequences in previous steps. After the privacy amplification, the final symmetric key is obtained at the transceivers, which can be applied to protect the data security.
\subsection{Related Works}
For PLKG, the key generation rate is a pivotal performance indicator to realize the ``one time pad" encryption. In general, the key generation rate mostly relies on the endogenous nature of the fading channel and cannot be guaranteed in harsh propagation environments. For example, in unmanned aerial vehicle (UAV) communication \cite{Gao9453771,8873597}, the high mobility of UAVs results in an extremely short channel coherence time, which entails formidable challenges to bidirectional channel probing and information reconciliation. For other harsh propagation environments, such as static indoor environments, negotiating a sufficiently random raw bit sequences is a laborious and time-consuming task due to the fact that the channel based attenuations are almost predictable \cite{9000831}. To address this problem, some previous works have tried to increase the randomness of the wireless fading channel by utilizing a single relay and/or cooperative relays \cite{6311224}. The approach of using relays can effectively increase the key generation rate, yet the participation
of untrusted relays can cause an information leakage of the secret key and the identification of the untrusted relays bring an additional overhead.

We now recall that a reconfigurable intelligent surface (RIS) is made of a planar digitally programmable metasurface, which can realize different wireless propagation functions by using a programmable field-programmable gate array (FPGA) as a smart controller \cite{CuiTieJunRIS}. In the last few years, RISs have been widely studied to improve the data transmission and wireless coverage \cite{9424177,9326394,8746155}.
Very recently, RISs are gradually investigated in the field of PLKG to increase the resilience of key generation with their excellent ability on the customization of channel environments \cite{ARXIV22GAO}. On one hand, the dynamic control of RIS on channel environments increases the wireless channel endogenous randomness, thereby improving the key generation rate in harsh propagation environments. On the other hand, the passive nature of RISs, that have no radio frequency chains, can help avoid the information leakage of the raw bit sequences, thereby relieving the issue of trustworthiness of relays. Furthermore, the plug-and-play scheme of RIS-assisted PLKG enhances the scalability and compatibility of the off-the-shelf devices.

Some existing works have utilized the electromagnetic wave control of RISs to improve the key generation rate. The RIS-assisted wireless communication security was originally discussed in reference \cite{LIASKOS20191}, where a programmable metasurface, namely HyperSurface, was developed to show groundbreaking security potential in indoor wireless communication. Most recently, to release the potential of RISs on improving the PLKG performance, RIS-assisted PLKG has been studied in some works \cite{9298937,ARXIV22,9569556}. In \cite{9298937}, the authors deployed a RIS near the user and utilized the statistical channel information to adjust its phase-shift, thereby increasing the achievable secret key capacity. One of the major challenges for PLKG is to guarantee the desirable key generation rate in low-entropy environments, i.e., static indoor environments \cite{ARXIV22GAO}. A wireless key generation architecture based on randomized channel responses from a RIS was proposed in \cite{9569556}, that it is generic for arbitrary devices and compatible with the existing PLKG implementations. The authors in \cite{9442833} proposed a novel
secret key generation scheme assisted by a RIS with discrete phase-shifts, where the dynamic time-varying channel environment is constructed by rapidly and randomly switching the phase of RIS elements. To verify the theoretical frameworks, the measurements of the actual performance of the RIS-assisted PLKG have been studied via proof-of-concept experiments and the practical design challenges have been discussed in \cite{ARXIV22GAO,9771319}. Different from arbitrary control, the RIS phase-shifts, active and passive beamforming, and transmit power, etc., have also been jointly optimized to maximize the key generation rate \cite{10106070,10001434,9771815}. For example, considering a general spatially correlated multi-antenna channel, the authors in \cite{10001434} proposed a joint transmit and reflective beamforming scheme to maximize the key generation rate. A RIS-assisted PLKG scheme was proposed in \cite{9771815}, which optimizes both the RIS reflecting coefficients and the transmit power to obtain a desirable key generation rate. Furthermore, the optimal RIS reflecting
coefficients for multiuser or multi-eavesdropper PLKG have also been investigated in \cite{10013094} and \cite{9663196}.

\subsection{Motivation and Contributions}
Nowadays, with the development of the 6G wireless networks, the endogenous security is becoming one of the critical concepts. Interestingly, by availing of the endogenous characteristics of wireless communications, i.e, channel state information (CSI), the air interface is given endogenous security capability. For example, by integrating the PLKG and data transmission, the data transmission can create the secret key for data encryption and in turn, the data encryption can enhance the trustworthiness of data transmission, which achieves the purpose of mutualism. Thus, the dual functions of communication and security can be mutually benefited by sharing spectrum, power and hardware resources, etc., achieving an ability of security and communications symbiosis. In our previous work, we have proposed the acronym integrated communications and security (ICAS) \cite{ARXIV22GAO}. By sharing the resource of RISs, an ICAS design paradigm that can simultaneously transmit and generate secret key with mutual benefit is possible. However, most of the related works ignore the design idea of this endogenous security, which can be an inefficient system design. Obviously, the RIS optimal configuration for PLKG may not be suitable for the optimal data transmission, and vice versa. The idea of ICAS system design is the main motivating factor behind this work. In this paper, we consider a novel RIS-assisted system that can simultaneously transmit and generate secret keys under the threats of a smart attacker. Specifically, the phase-shift of the RIS is configured to jointly optimize the data transmission rate and the key generation rate according to the malicious behaviors of the smart attacker. We formulate this attack-and-defense procedure as a secure transmission (ST) game based on game theory \cite{MJ1994,10121689}, and propose the optimal ST strategy in a static and dynamic game, respectively. By configuring the phase-shift as per the weight factor allocation, a mutual beneficial ``one time pad" communication can be achieved. Our main contributions are summarized as follows:
\begin{itemize}
  \item As one of the first works, by sharing the hardware resources of a RIS for two different tasks, we propose a novel ICAS design paradigm that can simultaneously guarantee the key performance indicators of transmitting and generating secret key to achieve a mutual beneficial ``one time pad" communication. In the proposed ICAS system, a smart attacker is considered, which can choose a sleeping or eavesdropping behavior based on the time-varying wireless environment.
  \item To maximize the data transmission rate and the key generation rate, we formulate the problem as a ST game and prove the existence of the Nash equilibrium (NE) point, and then derive the NE point of the static game with respect to the  malicious behavior and the optimal RIS phase-shift. For the dynamic game, we model the problem as a finite Markov decision process and propose a model-free reinforcement learning approach to obtain the NE point. Particularly, due to the fact that legitimate users cannot obtain the CSI of attacker in real-world, we develop a deep recurrent Q-network (DRQN) based dynamic ST strategy to configure the optimal RIS phase-shift with only partial observation information.
  \item The details of the proposed algorithm are discussed and its system complexity is analyzed. Simulations are conducted to evaluate the performance of the network convergence, the strategy gain and the RIS parameters, respectively. Our simulation results show that the proposed DRQN based dynamic ST strategy has a better performance than the benchmarks even with the partial observation information, and achieves the ``one time pad" communication by allocating a suitable weight factor for data transmission and PLKG.
\end{itemize}

\subsection{Overview}
The rest of this paper is organized as follows: The system model is presented in Section \ref{sec:2}. We derive the key generation rate and formulate the optimization problem in Section \ref{sec:3}. Section \ref{sec:4} discusses the optimal solutions. The simulation results are provided in Section \ref{sec:5} and the paper is concluded in Section \ref{sec:6}.

\section{system model and problem formulation}\label{sec:2}
In this section, we first give the channel model, and then illustrate the attack model. For ease of reference, Table \ref{Tab:notations} lists the important notations used in this paper.
\begin{table}[h]
\caption{List of important notations}
\label{Tab:notations}
\centering
\renewcommand\arraystretch{1.2}
\setlength{\tabcolsep}{6mm}{
\begin{tabular}{c|c}
  \hline
 \textbf{Symbols} & \textbf{Notations} \\
  \hline
  $D,d$ & Scalar \\
  $\mathbf{d}$ & Vector \\
  $\mathbf{D}$ &  Matrix \\
  $\mathbb{C}^{N\times1}$ & Dimension of the complex vector  \\
  $\mathbf{\Phi}$& Phase-shift matrix of RIS\\
  $\kappa$& Malicious behavior indicator\\
  $\mathcal{S},s$& State space, state\\
  $\mathcal{A},a$& Action space, action\\
  $r$& Immediate reward\\
  $J$& Size of state space\\
  $\eta$& Weight parameter of evaluated network\\
  $\eta^-$& Weight parameter of target network\\
  $\gamma$& Discount factor\\
  $\beta^t$& Learning rate in time step $t$\\
  $\mathcal{E}$& Episode of DRQN\\
  $T$& Time steps in each episode\\
  $Pr(\cdot|\cdot)$& Transition probability operator\\
  $[\cdot]^T$ & Transpose operator \\
  $\sum$ & Sum operator \\
  $\mathbb{E}[\cdot]$ & Expectation operator \\
  $\|\cdot\|$& Euclidian norm operator  \\
  $|\cdot|$& Size operator  \\
  \hline
\end{tabular}}
\end{table}
\begin{figure}[!ht]
 \centering
  \includegraphics[width=7.5cm]{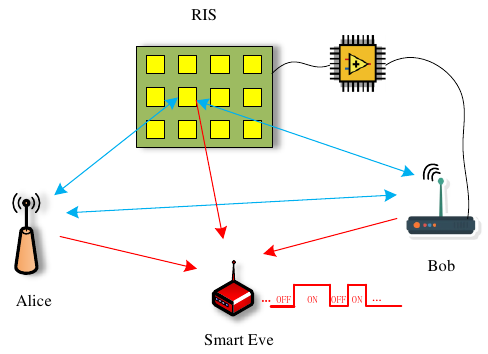}
  \caption{The schematic diagram of the three component RIS-assisted simultaneous transmission and generation of secret key in the presence of a smart attacker.}
  \label{Fig:BFL}
\end{figure}
\subsection{Channel Model}
We consider a static three component RIS-assisted ICAS system that consists of the legitimate transceivers Alice and Bob, and malicious user Eve. All of the participants are resource-limited devices with a single antenna. Alice and Bob intend to both transmit data and generate physical layer key, simultaneously, while the malicious Eve eavesdrops the key information over the wireless fading channel. In addition, a RIS is deployed in this network to improve both the communication and PLKG performances. In this case, the signal received at Alice can be denoted as
\begin{equation}
\label{eq:B2A}
y_{A}=\underbrace{\big(\mathbf{h}^T_{RA}\mathbf{\Phi}\mathbf{h}_{BR}
+h_{BA}\big)}_{h_{BRA}}x_B+n,
\end{equation}
where $\mathbf{h}_{BR}\in\mathbb{C}^{N\times1}$ is the channel gain from Bob to RIS, $\mathbf{h}_{RA}\in\mathbb{C}^{N\times1}$ is the channel gain vector from RIS to Alice, $h_{BA}\in\mathbb{C}$ is the channel gain of the direct link from Bob to Alice, $h_{BRA}$ is the equivalent channel, $x_B$ is the transmission signal of Bob and $n_A$ is the noise at Alice following zero-mean complex Gaussian distribution with variance $\sigma^2$. In particular, we consider the phase-shift of the RIS with finite resolution of $b$ bits, where the $n$th reflection element of the RIS can be denoted as $\theta_n=\{\frac{2\pi p}{2^b}|0\leq p\leq2^b-1,p\in \mathbb{Z}\}$ \cite{8746155}. Then, the diagonal phase-shift matrix of RIS with $N$ reflection elements is represented by $\mathbf{\Phi}=\text{diag}\big[\alpha_1e^{j\theta_1},\alpha_2e^{j\theta_2},\ldots,\alpha_Ne^{j\theta_N}\big]$ with the amplitude $\alpha_n=1, n\in\{1,\ldots,N\}$. As all links are reciprocal, the received signal from Alice to Bob can be represented by
\begin{equation}
\label{eq:A2B}
y_{B}=\underbrace{\big(\mathbf{h}^T_{RB}\mathbf{\Phi}\mathbf{h}_{AR}
+h_{AB}\big)}_{h_{ARB}}x_A+n,
\end{equation}
where $x_A$ is the transmission signal of Alice. We focus on a narrowband fading channel, where the channel gains $\mathbf{h}_{RA},\mathbf{h}_{BR}, \mathbf{h}_{RB}$, $\mathbf{h}_{AR}$ and the direct link channel $h_{BA},h_{AB}$ are independent and identically distributed complex Gaussian random vector (variables) having zero-mean and unit variance entries.
Similarly, the received signal of Eve from Bob or Alice can be written by exchanging the subscripts $A$ in \eqref{eq:B2A} or $B$ in \eqref{eq:A2B} with $E$, respectively. For example, the
received signal from Alice to Eve is given by
\begin{equation}
\label{eq:A2E}
y_{E}=\underbrace{\big(\mathbf{h}^T_{RE}\mathbf{\Phi}\mathbf{h}_{AR}
+h_{AE}\big)}_{h_{ARE}}x_A+n.
\end{equation}
\subsection{Attack Model}
We consider that the malicious Eve is a smart attacker who has two types of behaviors, including a sleeping behavior and eavesdropping behavior. We refer to this as a smart eavesdropping attack and define these behaviors as $\kappa=1$ for eavesdropping and $\kappa=0$ for sleeping. Specifically, the behaviors of Eve are executed based on the observation of the time-varying wireless environment with the Bayesian inference. For instance, the malicious Eve can keep silent to save resources when the quality of the wiretap channel is inferred to be bad. On the contrary, Eve chooses to monitor the process of PLKG over the wireless fading channel when the quality of the wiretap channel is inferred to be good. The transition probability of malicious behavior in time step $t$ can be denoted as
\begin{equation}
\label{eq:trans}
Pr(\kappa^t|\kappa^{t-1})=Pr(h^{t}_{ARB},h^{t}_{ARE}|h^{t-1}_{ARB},h^{t-1}_{ARE}).
\end{equation}
Furthermore, the PLKG protocol is assumed to be public to both the legitimate transceivers and adversary. We point out that the transition probability of malicious behavior is assumed to be unknown to Alice and Bob.

\section{Key Generation Rate and Problem Formulation}\label{sec:3}
In this section, we first provide the key generation procedure, and then derive the closed-form expression of the key generation rate with the assistance of RIS. After that, we formulate the optimization problem.
\subsection{Key Generation Procedure}
Compared with RSS, the CSI is a fine grained quantity, which can provide more refined features for PLKG. Thus, we use the CSI as the endogenous random source of key generation. The key generation procedure for RIS-assisted ICAS system includes the following steps.

First, Alice sends the pilot signal $x^p_A$ to Bob, and Bob probes the CSI. In this case, the received probing signal at Bob is given by
\begin{align}
y^p_{B}=(\mathbf{h}^T_{RB}\mathbf{\Phi}\mathbf{h}_{AR}
+h_{AB})x^p_A+n.
\end{align}
Then, Bob obtains the estimated CSI via the least
squares (LS) channel estimation, which is denoted as
\begin{align}
\hat{h}_{ARB}=y^p_{B}x^{p*}_A=\mathbf{h}^T_{RB}\mathbf{\Phi}\mathbf{h}_{AR}
+h_{AB}+nx^{p*}_A,
\end{align}
where the pilot signal satisfies $x^{p}_Ax^{p*}_A=1$ and $nx_A^{p*}$ is the channel estimation error.

Second, for the reverse channel probing from Bob to Alice, Alice can obtain the CSI estimate by calculating the pilot signal $x^{p}_B$, which is given by\footnote{In this case, the key generation at Alice can be executed via wireless control channel, i.e., CSI feedback, acknowledgement (ACK), etc.}
\begin{align}
\hat{h}_{BRA}=\mathbf{h}^T_{RA}\mathbf{\Phi}\mathbf{h}_{BR}
+h_{BA}+nx^{p*}_B.
\end{align}

The next step is to quantize the channel estimation, i.e., $\hat{h}_{ARB}$ and $\hat{h}_{BRA}$ into raw bit sequences, i.e., $\mathbf{q}_A$ and $\mathbf{q}_B$, based on the bit quantization methods \cite{7539590}. Due to the influence of estimation error, noise, and asymmetry of radio frequency fingerprint, etc, the raw bit sequences between Alice and Bob are inconsistent. Then, Alice and Bob conduct the information reconciliation and privacy amplification steps based on the raw bit sequences to generate the final secure bits. The information reconciliation and privacy amplification are, thus, not particularly studied, and we refer the interested readers to the related literature (see for example \cite{7448884} and references therein).

\subsection{Key Generation Rate}
Given the channel reciprocity of all links in time division duplex (TDD) systems, the key generation rate from Alice to Bob, and Bob to Alice are equal. Therefore, we assume that Bob is the active initiator of key generation and data transmission, thus, taking the key generation rate at Alice side as a case. Due to the loss of bit quantization, we have the mutual information $I(\hat{h}_{ARB};\hat{h}_{BRA})\geq I(\mathbf{q}_A;\mathbf{q}_B)$. In this case, the key generation rate can be characterized by a upper bound, which is the mutual information of CSI given the observation of Eve. When $\kappa=1$, which indicates the behavior of Eve is eavesdropping, the key generation rate can be expressed as \cite{9627160}
\begin{align}
\label{eq:MI}
\mathcal{R}_k^1&=\frac{1}{T_s}I(\hat{h}_{ARB};\hat{h}_{BRA}|\hat{h}_{ARE})~~~~~~~~~\nonumber\\
&=\frac{1}{T_s}\log_2\frac{\det(\mathbf{R}_{AE})\det(\mathbf{R}_{BE})}{\det(\mathbf{R}_E)\det(\mathbf{R}_{ABE})}~\text{(bit/s)},
\end{align}
where $T_s$ is the time for each channel estimation round, $\det(\cdot)$ is the matrix determinant, while $\mathbf{R}_{AE}$, $\mathbf{R}_{BE}$, $\mathbf{R}_E$ and $\mathbf{R}_{ABE}$ are the covariance matrixes. When $\kappa=0$, Eve is in the sleep state, and the key generation rate can be given by
\begin{align}
\label{eq:MIS}
\mathcal{R}_k^0&=\frac{1}{T_s}I(\hat{h}_{ARB};\hat{h}_{BRA})\nonumber\\
&=\frac{1}{T_s}\log_2\frac{\det(\mathbf{R}_{A})\det(\mathbf{R}_{B})}{\det(\mathbf{R}_{AB})}~\text{(bit/s)}.
\end{align}
\begin{proposition}
\label{eq:proposition}
For a RIS-assisted SISO wireless communication system, the cross-correlation of the CSI between the transmitter $\tilde{t}$ and two receivers $r1,r2$, i.e., $h_{\tilde{t}Rr1}$ and $h_{\tilde{t}Rr2}$, is given by
\begin{align}
\label{eq:CCor}
R_{r1,r2}=\mathbf{v}^H_\theta\mathbf{R}_{r1,r2}\mathbf{v}_\theta+R^D_{r1,r2},
\end{align}
while the auto-correlations between the transmitter $\tilde{t}$ and receiver $r1$, $r2$ are given by
\end{proposition}
\begin{align}
\label{eq:ACor}
R_{r1}=\mathbf{v}^H_\theta\mathbf{R}_{r1}\mathbf{v}_\theta+R^D_{r1},\nonumber\\
R_{r2}=\mathbf{v}^H_\theta\mathbf{R}_{r2}\mathbf{v}_\theta+R^D_{r2},
\end{align}
where the vector $\mathbf{v}_\theta=[\alpha_1e^{j\theta_1},\alpha_2e^{j\theta_2},\ldots,\alpha_Ne^{j\theta_N}]^T$, $\mathbf{R}_{r1,r2}$ is the cross-correlation matrix of the RIS links and $R^D_{r1,r2}$ is the cross-correlation of the direct link. Moreover, $\mathbf{R}_{r1}$ and $\mathbf{R}_{r2}$ are the auto-correlation matrices of the RIS links, while $R^D_{r1}$, $R^D_{r2}$ are the auto-correlations of the direct links, respectively.
\begin{IEEEproof}
See Appendix \ref{sec:AP1}.
\end{IEEEproof}
Based on Proposition \ref{eq:proposition}, we can rewrite Eq. (\ref{eq:MI}) as
\begin{align}
\mathcal{R}_k^1=\frac{1}{T_s}\big[2\log_2\det(\mathbf{R}_{AE})-\log\det(\mathbf{R}_E)-\log\det(\mathbf{R}_{ABE})\big],
\end{align}
where
\begin{align}
\det(\mathbf{R}_{AE})&=(R_A+\sigma^2)(R_E+\sigma^2)-|R_{A,E}|^2,\nonumber\\
\det(\mathbf{R}_{E})&=R_E+\sigma^2,\nonumber\\
\det(\mathbf{R}_{ABE})&=
(2R_A\sigma^2+\sigma^4)(R_E+\sigma^2)-2\sigma^2|R_{A,E}|^2.~\nonumber
\end{align}
Similarly, we can rewrite Eq. (\ref{eq:MIS}) as
\begin{align}
\mathcal{R}_k^0=\frac{1}{T_s}\big[2\log_2\det(\mathbf{R}_{A})-\log\det(\mathbf{R}_{AB})\big],
\end{align}
where
\begin{align}
\det(\mathbf{R}_{A})&=(R_A+\sigma^2),~~~~~~~~~~~\nonumber\\
\det(\mathbf{R}_{AB})&=(2R_A\sigma^2+\sigma^4).~~~~~\nonumber
\end{align}
Without loss of generality, we normalize the variance of the noise to $\sigma^2=1$ for simplicity and use the symbol $h$, i.e., $h_{ARB}$, uniformly as the CSI observation. According to (\ref{eq:CCor}) and (\ref{eq:ACor}), the key generation rate is given in (\ref{eq:Krate}), (\ref{eq:Krate1}) and the leaked key generation rate at Eve is given in (\ref{eq:KrateE}), which can be found at the top of the next page.

\subsection{Problem Formulation}
\begin{figure}[!ht]
 \centering
  \includegraphics[width=8cm]{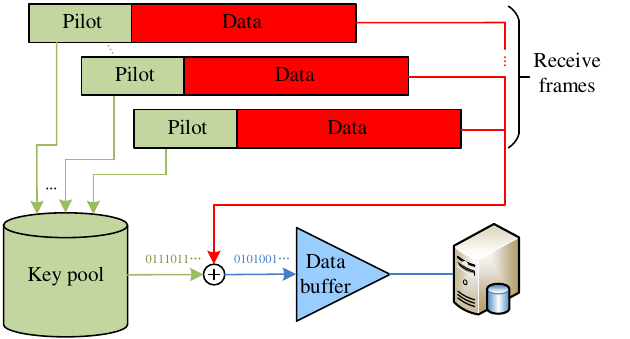}
  \caption{The schematic diagram of the simultaneously
transmitting and generating secret key at the receiver side. The secret key is generated from CSI estimation via pilot and stored in the key pool. The receive data is decrypted by the secret key of the key pool in real time via an XOR operation.}
  \label{Fig:frame}
\end{figure}
\begin{figure*}
\begin{align}
\label{eq:Krate}
\mathcal{R}^1_k=\frac{1}{T_s}\log_2\bigg(\frac{\big((\mathbf{v}^H_\theta\mathbf{R}_{A}\mathbf{v}_\theta+R^D_{A}+1)(\mathbf{v}^H_\theta\mathbf{R}_{E}\mathbf{v}_\theta+R^D_{E}+1)-|\mathbf{v}^H_\theta\mathbf{R}_{A,E}\mathbf{v}_\theta+R^D_{A,E}|^2\big)^2}{(\mathbf{v}^H_\theta\mathbf{R}_{E}\mathbf{v}_\theta+R^D_{E}+1)\big((2\mathbf{v}^H_\theta\mathbf{R}_{A}\mathbf{v}_\theta+2R^D_{A}+1)(\mathbf{v}^H_\theta\mathbf{R}_{E}\mathbf{v}_\theta+R^D_{E}+1)-2|\mathbf{v}^H_\theta\mathbf{R}_{A,E}\mathbf{v}_\theta+R^D_{A,E}|^2\big)}\bigg),\\
\label{eq:Krate1}
\mathcal{R}^0_k=\frac{1}{T_s}\log_2\bigg(\frac{(\mathbf{v}^H_\theta\mathbf{R}_{A}\mathbf{v}_\theta+R^D_{A}+1)^2}{2\mathbf{v}^H_\theta\mathbf{R}_{A}\mathbf{v}_\theta+2R^D_{A}+1}\bigg),~~~~~~~~~~~~~~~~~~~~~~~~~~~~~~~~~~~~~~~~~~~~~~~~~~~~~~~~~~~~~~~~~~~~~~~~~~~~~~~~~~~~~~~~\\
\label{eq:KrateE}
\mathcal{R}_{k,E}=\frac{1}{T_s}\log_2\bigg(\frac{(\mathbf{v}^H_\theta\mathbf{R}_{A}\mathbf{v}_\theta+R^D_{A}+1)(\mathbf{v}^H_\theta\mathbf{R}_{E}\mathbf{v}_\theta+R^D_{E}+1)}{\big((\mathbf{v}^H_\theta\mathbf{R}_{A}\mathbf{v}_\theta+R^D_{A}+1)(\mathbf{v}^H_\theta\mathbf{R}_{E}\mathbf{v}_\theta+R^D_{E}+1)-|\mathbf{v}^H_\theta\mathbf{R}_{A,E}\mathbf{v}_\theta+R^D_{A,E}|^2\big)^2}\bigg).~~~~~~~~~~~~~~~~~~~~~~~~~~~
\end{align}
\hrulefill
\end{figure*}

According to the derived key generation rate, we can formulate the optimization problem as follows. The considered RIS-assisted ICAS system, simultaneously transmits and generates a secret key, which is shown in Fig. \ref{Fig:frame}.
On one hand, we consider to maximize the upper bound of the data transmission rate $\mathcal{R}_d$ to maintain consistency of the indicator with the key generation rate. Due to the reciprocity of all links, the data transmission rate at Alice can be denoted as
\begin{eqnarray}
\label{eq:regodic}
\mathcal{R}_d=B\log_2\big(1+\mathbb{E}\{|\mathbf{v}^T_\theta\mathbf{\Lambda}_{RA}\mathbf{h}_{BR}
+h_{BA}|^2\}\big)~\text{(bit/s)},
\end{eqnarray}
where $B$ is the available bandwidth of the transmit signal.
On the other hand, the transceivers are to maximize the key generation rate $\mathcal{R}_k$ with the potential eavesdropping threat of a smart Eve. To this end, we optimize the phase-shift vector $\mathbf{v}_\theta$ of the RIS to maximize the dual performances of the ICAS system. The optimization problem can be written as
\begin{eqnarray}
\label{eq:PF}
\mathbb{P}:&&\max_{\mathbf{v}_\theta}~~~w_d\mathcal{R}_d+w_k\mathcal{R}^\kappa_k,\\
\label{eq:jiao}
&&\text{s.t.}~~~~~~ 0\leq\theta_n\leq2\pi,~\forall n\in\{1,\ldots,N\},\\
\label{eq:ka}
&&~~~~~~~~~\kappa\in\{0,1\},
\end{eqnarray}
where $\mathcal{R}^\kappa_k$ is a unified representation of $\mathcal{R}^\kappa_1$, $\mathcal{R}^\kappa_0$, and can be found in (\ref{eq:KrateCom}) at the top of the next page, and $w_d,w_k=1-w_d\in[0,1]$ are the weight factors balancing the importance of transmitting and generating secret key. The constraint \eqref{eq:jiao} guarantees that the phase-shift range of each reflection element is $[0,2\pi]$ and the constraint \eqref{eq:ka} shows that the smart Eve has two behaviors to choose from. In our system model, the RIS-assisted communication in the presence of a smart attacker is a typical attack-and-defense procedure, which can be mathematically expressed via game theory. Thus, we transform the optimization problem into a non-cooperative game, which we refer to as a ST game. In the formulated ST game, Alice chooses its own best strategy to mitigate the attack of Eve, whilst the smart Eve chooses its malicious behavior to balance its benefits, such as saving resources or eavesdropping the key information. Specifically, the ST game can be defined as a tuple $\mathbb{G}\langle\{A,E\},\{\mathbf{v}_\theta,\kappa\},\{u_A,u_E\}\rangle$, where
\begin{figure*}
\begin{align}
\label{eq:KrateCom}
\mathcal{R}^\kappa_k=\frac{1}{T_s}\log_2\bigg(\frac{\big((\mathbf{v}^H_\theta\mathbf{R}_{A}\mathbf{v}_\theta+R^D_{A}+1)(\mathbf{v}^H_\theta\mathbf{R}_{E}\mathbf{v}_\theta+R^D_{E}+1)^\kappa-\kappa|\mathbf{v}^H_\theta\mathbf{R}_{A,E}\mathbf{v}_\theta+R^D_{A,E}|^2\big)^2}{(\mathbf{v}^H_\theta\mathbf{R}_{E}\mathbf{v}_\theta+R^D_{E}+1)^\kappa\big((2\mathbf{v}^H_\theta\mathbf{R}_{A}\mathbf{v}_\theta+2R^D_{A}+1)(\mathbf{v}^H_\theta\mathbf{R}_{E}\mathbf{v}_\theta+R^D_{E}+1)^\kappa-2\kappa|\mathbf{v}^H_\theta\mathbf{R}_{A,E}\mathbf{v}_\theta+R^D_{A,E}|^2\big)}\bigg).
\end{align}
\hrulefill
\end{figure*}
\begin{eqnarray}
\label{eq:PF}
&&u_A(\mathbf{v}_\theta,\kappa)=w_d\mathcal{R}_d+w_k\mathcal{R}^\kappa_k,\\
&&u_E(\mathbf{v}_\theta,\kappa)=\kappa(\mathcal{R}_{k,E}-C_E),
\end{eqnarray}
is the utility function with $C_E$ being the cost for wireless channel eavesdropping.

\section{Proposed Solutions}\label{sec:4}
In this section, we derive the NE point existence of the ST game and then study a case of the static game which corresponds to the snapshot of the ST game. Then, corresponding to the whole evolutionary process of the ST game, we consider the dynamic ST game and propose a DRQN based dynamic ST strategy, thereby obtaining the RIS optimization configuration for simultaneously transmitting and generating secret key.

\subsection{NE of ST Game}
\begin{definition}
Let $\Pi(\mathcal{A}_i)$ be the set of all probability distributions over $\mathcal{A}_i$, while the set of mixed strategy for participant $i$ can be denoted as $\mathcal{P}_i=\Pi(\mathcal{A}_i)$. Then, the set of mixed strategy profiles is denoted as the Cartesian product of the individual mixed strategy sets, $\mathcal{P}=\mathcal{P}_1\times\ldots\times \mathcal{P}_i$.
\end{definition}
For the formulated ST game $\mathbb{G}$, a mixed strategy is used for the game participants. Thus, the expected utility of the mixed strategy profile $P(\mathbf{v}_\theta,\kappa)\in \mathcal{P}$ for the participants Alice and Eve can be written as
\begin{eqnarray}
\label{eq:defNE}
&&\bar{u}_A(\mathbf{v}_\theta,\kappa)=\sum_{ \mathcal{A}_{A}\times\mathcal{A}_E}u_A(a)\prod_{i\in\{A,E\}}P_i(a_i)\nonumber\\
&&\bar{u}_E(\mathbf{v}_\theta,\kappa)=\sum_{\mathcal{A}_{A}\times\mathcal{A}_E}u_E(a)\prod_{i\in\{A,E\}}P_i(a_i).
\end{eqnarray}
To investigate the game from an individual participant's point of view, this leads to the most influential solution concept, i.e., Nash equilibrium. The Nash equilibrium point is one of the most important concepts
in a game, which can be defined as follows \cite{MJ1994}.
\begin{definition}
\label{eq:NE}
For the ST game $\mathbb{G}$, the mixed strategy profile $(\mathbf{v}_\theta^\star,\kappa^\star)$ is represented as the NE point, if and only if
\begin{eqnarray}
\label{eq:defNE}
&&\bar{u}_A(\mathbf{v}_\theta^\star,\kappa^\star)\geq \bar{u}_A(\mathbf{v}_\theta,\kappa^\star),~0\leq\theta_n\leq2\pi\nonumber\\
&&\bar{u}_E(\mathbf{v}_\theta^\star,\kappa^\star)\geq \bar{u}_E(\mathbf{v}_\theta^\star,\kappa),
~\forall \kappa\in\{0,1\}.
\end{eqnarray}
\end{definition}
The above equation presents the convergence conditions of the NE point for the ST game, where neither Alice or Eve have the motivation to break the balance in the game. To validate the existence of a Nash
equilibrium in our formulated game, we give the following proposition.
\begin{proposition}
The formulated ST game with a finite number of participants and action profiles has at least one Nash equilibrium.
\end{proposition}
\begin{IEEEproof}
See Appendix \ref{sec:AP2}.
\end{IEEEproof}
Next, we define the duration of one type of malicious behavior as one time step. Then, for one time step, the formulated ST game can be regarded as a static game, where a NE point for the static game is given as follows.
\begin{proposition}
There is a NE point $(\mathbf{v}_\theta^\star,0)$, where $\mathbf{v}_\theta^\star$ satisfies
\begin{eqnarray}
\mathbf{v}^{\star H}_\theta \mathbf{R}_A\mathbf{v}^\star_\theta=N\lambda_{\max}(\mathbf{R}_A).
\end{eqnarray}
The following condition must hold
\begin{eqnarray}
C_E\geq \mathcal{R}_{k,E}|_{\mathbf{v}_\theta=\mathbf{v}_\theta^\star}.
\end{eqnarray}
\end{proposition}
\begin{IEEEproof}
See Appendix \ref{sec:AP3}.
\end{IEEEproof}
\begin{remark}
We cannot attain a specific NE point in the formulated ST game, except for a static ST game. At the NE point $(\mathbf{v}_\theta^\star,0)$, Eve is in the sleeping state which means there is no eavesdropping behavior in the RIS-assisted ICAS system, which is a pure strategy. Specifically, the optimal $\mathbf{v}_\theta^\star$ indicates that all the RIS phase-shifts should be aligned with the direction of the maximum eigenvalue of the RIS link auto-correlation matrix, which can bring a $N$ order gain. Interestingly, in this static game, and given the monotonicity of the optimization objective with respect to the channel auto-correlation, the optimal $\mathbf{v}_\theta^\star$ can simultaneously maximize the data transmission rate and the key generation rate, no matter what the weight factors is. That is, in this static game, separately optimizing thr RIS phase-shift for the data transmission rate or the key generation rate can also achieve the same mutualistic purpose as the joint optimization.
\end{remark}

\subsection{DRQN for Dynamic Game}
\begin{figure}[!ht]
 \centering
  \includegraphics[width=7.2cm]{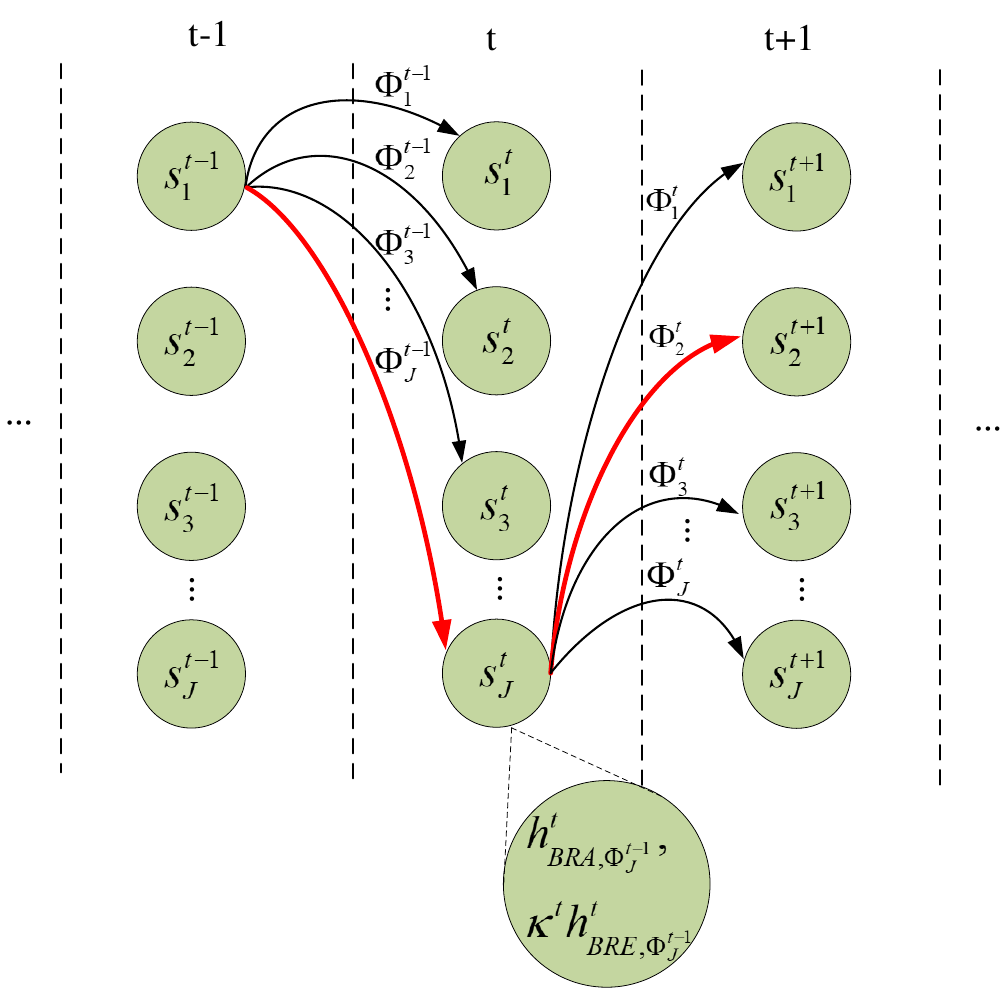}
  \caption{Schematic diagram of the Markov decision process for RIS-assisted simultaneous transmission and generation of secret key.}
  \label{Fig:MDP}
\end{figure}

Considering the dynamic wireless environments and the dynamic malicious behavior of smart Eve, the whole evolutionary process of the ST game is a dynamic game. On one hand, from \eqref{eq:trans}, we can find that the transition probability of malicious behavior $Pr(\kappa^t|\kappa^{t-1})$ is equal to the transition probability between the current CSI observation $h^{t}_{BRA},h^{t}_{BRE}$ and the previous CSI observation $h^{t-1}_{BRA},h^{t-1}_{BRE}$. Thus, the transition probability of malicious behavior has a Markov property and the smart attack by Eve is usually represented by a two-state Markov decision process \cite{GAO}. On the other hand, from the perspective of the legitimate user, Alice observes the current malicious behavior $\kappa^t$, and the CSI $h^{t}_{BRA},h^{t}_{BRE}$ which are a function of the previous RIS phase-shift $\mathbf{\Phi}^{t-1}$. Next, Alice chooses a RIS phase-shift $\mathbf{\Phi}^t$ to feedback the environment observation. Then, after configuring the current RIS phase-shift $\mathbf{\Phi}^t$, the CSI transfers from $h^{t}_{BRA},h^{t}_{BRE}$ to $h^{t+1}_{BRA},h^{t+1}_{BRE}$. Therefore, it can be found that the action change of Alice produces a finite state Markov decision process, which is illustrated in Fig. \ref{Fig:MDP}. For example, the state transfers from $s_1^{t-1}$ to $s_J^t$ after taking the RIS phase-shift $\mathbf{\Phi}_J^{t-1}$ in time step $t-1$, and from $s_J^t$ to $s_2^{t+1}$ after taking the RIS phase-shift $\mathbf{\Phi}_2^t$ in time step $t$, which is represented as the red state transfer path.

It has been shown that reinforcement learning is a powerful tool to solve the NE point of the dynamic game which satisfies a Markov decision process \cite{8873597,8714026}. For example, reinforcement learning has been utilized in different RIS-assisted physical cross-layer security scenarios to optimize the RIS phase-shift in time series \cite{10130368,9462487,9206080,9961877}, such as anti-jamming, secure communication, etc. Therefore, we herein adopt reinforcement learning to design the optimal dynamic ST strategy. Since the legitimate user Alice cannot know the transition probability of the finite state Markov decision process in time, we use a model-free reinforcement learning such as Q-learning to optimize the ST strategy of RIS phase-shift to maximize the data transmission rate and key generation rate, simultaneously. By regarding Alice as an intelligent agent, in each time step, Alice perceives its surrounding dynamic environment and adjusts the RIS phase-shift based on the environment feedback. We define a 3-tuple $\langle \mathcal{S}, \mathcal{A}, r\rangle$ to characterize such interaction loop and omit the subscript $A$ in $\mathcal{A}_A$ for simplicity of expression. The details are characterized as follows:

\textsf{Action:} The action of Alice for time step $t$ can be written as $a^t$, which corresponds to the optimization parameters, including the reflection elements of $\mathbf{\Phi}$. Thus, an action of Alice for time step $t$ can be written as $a^t=\{\mathbf{\Phi}^t\}, a\in \mathcal{A}$.

\textsf{State:} The wireless environment interaction among Alice, Bob and Eve is represented by state $s$. Based on the analysis of Fig. \ref{Fig:MDP}, in time step $t$, Alice observes the current environment information with respect to the CSI of the game participants, the malicious behavior $\kappa^{t}$ and the action of the previous time step, which are represented as the state $s^t$. Thus, the state of Alice for time step $t$ can be given by
\begin{align}
s^t=\{\mathbf{h}_{BR}^{t},\mathbf{h}_{RA}^{t},\mathbf{h}_{RE}^{t},h^{t}_{BA},h^{t}_{BE},\kappa^{t},\mathbf{\Phi}^{t-1}\}\nonumber\\
=\{h^t_{BAR,\Phi^{t-1}},\kappa^th^t_{BAE,\Phi^{t-1}}\},~~~~~~~~~~~~~
\end{align}
where $s\in \mathcal{S}$.

\textsf{Reward:} The immediate reward of the dynamic game for Alice in any time step is represented by $r$, which is set to be proportional to the weighted sum of the data transmission rate and key generation rate. In this case, the reward function of Alice is given by
\begin{align}
\label{eq:Rm}
r=w_d\mathcal{R}_d+w_k\mathcal{R}^\kappa_k,
\end{align}
where $\mathcal{R}^\kappa_k$ in \eqref{eq:KrateCom} is the combination of \eqref{eq:Krate} and \eqref{eq:Krate1}, shown at the top of the previous page.
\begin{remark}
From the state space, action space and reward, we can find that the receiver only needs to know the CSI of the equivalent channel related to $h^t_{BRA}$ and $h^t_{BRE}$. Thus, in our system, there is no need to consider the CSI acquisition problem of the RIS segmented channel.
\end{remark}
The goal of the Q-learning is to find an optimal policy $\pi$ that maximizes the value function which is defined as $V(s)=\mathbb{E}_\pi[\Sigma_{\tau=0}^\infty\gamma^{\tau}r_{t+\tau+1}|s^t=s]$ with the discount factor $\gamma$. Specifically, the agent performs exploration and exploitation with the $\varepsilon$-greedy policy to determine the policy $\pi$, which maps the states to the actions, i.e., $\pi:\mathcal{S}\rightarrow \mathcal{A}$. The Q-function can be used to evaluate the performance of a policy that represents $V(s)$ following the policy $\pi$ by performing action $a$ in state $s$. That is, $Q(s,a)=\mathbb{E}_\pi[\Sigma_{\tau=0}^\infty\gamma^{\tau}r_{t+\tau+1}|s^t=s,a^t=a]$ and the one step ahead Q-function is updated by
\begin{eqnarray}
Q(s,a)\leftarrow Q(s,a)+\beta^t\big[r_t+\gamma\max_{a'} Q(s',a')-Q(s,a)\big],
\end{eqnarray}
where $\beta^t$ is the learning rate and $a'$ is the action selected by the policy on the next state $s'$. Here, we use a convolutional neural network (CNN) to approximate the action-value function based on the following two considerations.

On one hand, when the state space is large, Q-learning can suffer the curse of high dimensionality. On the other hand, since the smart attack of Eve is a passive attack, the legitimate users Alice and Bob cannot obtain the CSI of Eve in the real-world, i.e., $\mathbf{h}_{RE}$ and $h_{BE}$. Therefore, Q-learning cannot work well with only a partial observation state. To deal with these problems, we model the dynamic ST game as a partially observation Markov decision process and
 propose a deep recurrent Q-network (DRQN) based dynamic ST strategy, which combines Q-learning with CNN and long short term memory (LSTM) for the optimal policy learning.
\begin{figure*}[!ht]
 \centering
  \includegraphics[width=15.2cm]{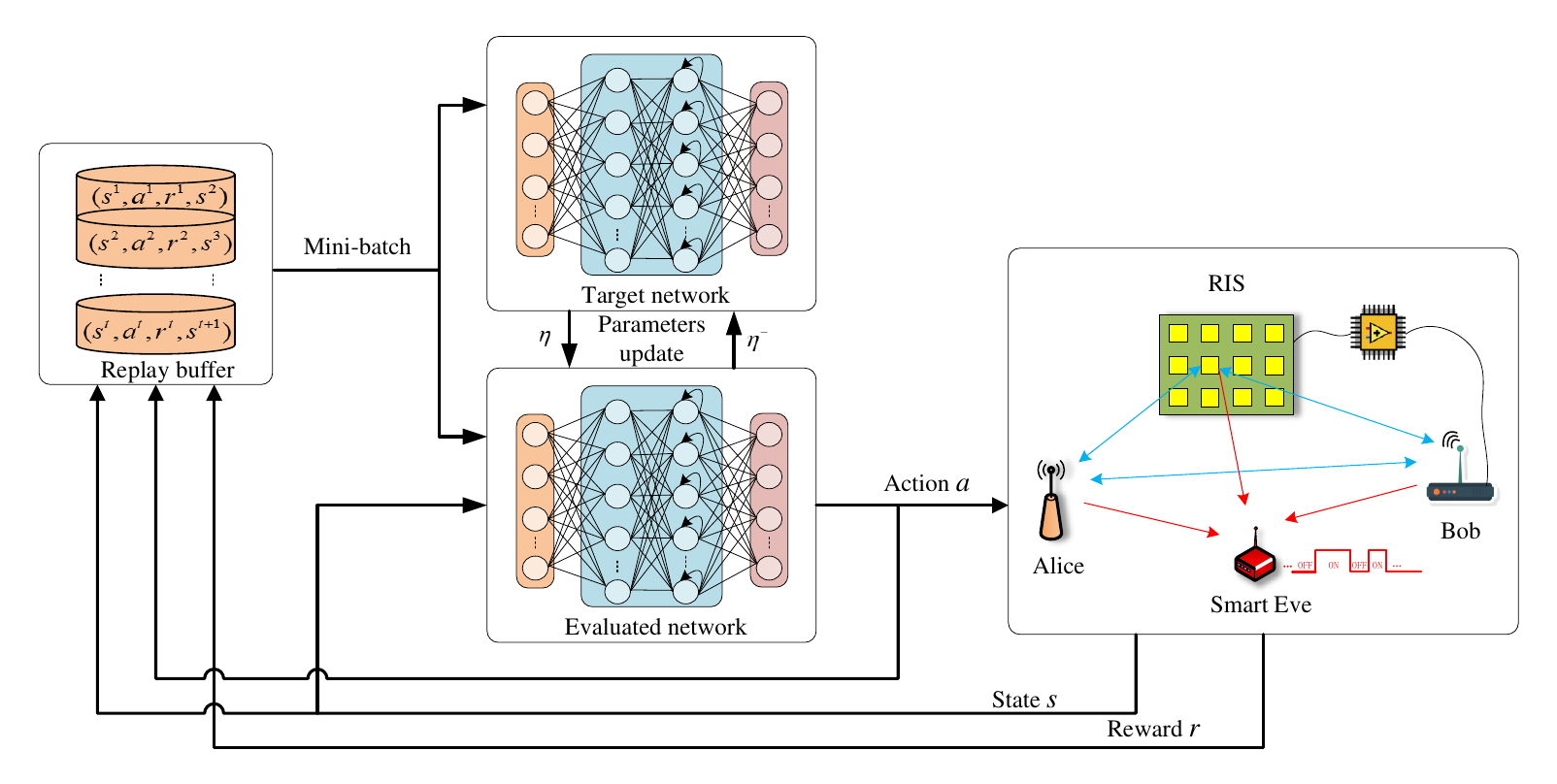}
  \caption{Illustration of the DRQN based dynamic ST strategy for the RIS-assisted ICAS system in the presence of a smart attacker.}
  \label{Fig:BFL}
\end{figure*}
In the training stage, we train a DRQN to approximate the optimal policy with the observation state $s\in\mathcal{S}$. Then, in the deployment stage, we apply the pre-trained DRQN to perform the optimal policy with only a partial observation state, i.e., $o=s/\{\mathbf{h}_{RE},h_{BE},\kappa\}$. Specifically, the framework of the training stage for DRQN is shown in Fig. \ref{Fig:BFL}, including the evaluated Q-network and the target Q-network with the LSTM layer to choose the possible phase-shift of the RIS. First, by observing the environment state $s^t$ at time step $t$, the state is preprocessed into a matrix. Then, the Q-value with respect to the action space $\mathcal{A}$ is calculated via the evaluated network. The agent performs the action $a^t$ with the $\varepsilon$-greedy policy and obtains the reward $r^t$ according to \eqref{eq:Rm}. After that, the state is changed from $s^t$ to $s^{t+1}$. Therein, a replay memory $\mathcal{D}$ is adopted to store the experience of Alice, where the experience $e(s^t,a^t,r^t,s^{t+1})$ records the transition among state, action and reward in time step $t$. To reduce the complexity, we use the bootstrapped random update strategy \cite{hausknecht2015deep}. With a randomly minibatch sample from replay memory with size $|\mathbf{e}|$, the weight parameter $\eta$ can be updated via a stochastic gradient descent (SGD). The loss function of the weight parameter $\eta$ is given by
\begin{eqnarray}
\label{eq:criticU}
\mathcal{L}(\eta)=\frac{1}{|\mathbf{e}|}\sum_{|\mathbf{e}|}|Y^{t=e}-Q(s^{t=e},a^{t=e};\eta)|^2,
\end{eqnarray}
where
\begin{align}
\label{eq:yt}
Y^t=\left\{
  \begin{array}{ll}
    r^t+\gamma\max Q(s',a',\eta^-), & \hbox{ $t<|\mathbf{e}|$}, \\
    r^t, & \hbox{ $t=|\mathbf{e}|$},
  \end{array}
\right.
\end{align}
with $\eta^-$ being the weight parameter of the target network.
The policy gradient of the loss function with respect to the weight parameter $\eta$ is given by
\begin{align}
\label{eq:decent}
\nabla\mathcal{L}(\eta)\approx
\frac{\sum_{|\mathbf{e}|}(Y^{t=e}-Q(s^{t=e},a^{t=e};\eta)\nabla_{\eta}Q(s^{t=e},a^{t=e};\eta))}{|\mathbf{e}|}.
\end{align}
The weight parameter of the evaluated network is updated by
$\eta\leftarrow\eta+\mu\nabla\mathcal{L}(\eta)$
and the weight parameter of the target network is updated by $\eta^-\leftarrow\eta^-+\mu\nabla\mathcal{L}(\eta^-)$ in every $S$ steps,
respectively, where $\mu\ll1$ is a small constant. When the DRQN is trained, we can deploy the evaluated network on Alice or Bob side to conduct the phase-shift control of the RIS. In this stage, the legitimate users Alice and Bob cannot obtain the CSI of Eve, thus, they only utilize the partial observation
state $o$ to perform the control policy of RIS phase-shift. The details of the proposed DRQN based dynamic ST strategy can be summarized as Algorithm \ref{alg:1}, which includes the training stage and the deployment stage.
\begin{remark}
The partial observation state $o$ is the lossy information of the observation state $s$, which lacks the CSI and malicious behavior information of Eve. Fortunately, the CSI sequences of the legitimate channel and the eavesdropping channel are spatially correlated, while the malicious behavior information and the CSI observation are temporally correlated. Thus, the LSTM layer can predict the loss information of the partial observation state $o$ by learning the context relation of the observation state $s$ in time sequence, and then handle the loss information of the partial observation state $o$ via the recurrent control.
\end{remark}
\begin{algorithm}[!h]
\caption{DRQN based dynamic ST strategy training}\label{alg:1}
\begin{algorithmic}[1]
\STATE \textbf{Initialization:} Initialize the RIS in the appointed two-dimensional space and set the phase-shift of RIS, randomly. Let the neural network parameters of the evaluated network $\eta_0$ and the target network $\eta_0^-$ as $\eta_0=\eta_0^-$.
\STATE \textbf{\underline{TRAINING STAGE}:}
\FOR {each training episode}
\STATE Select an initial state $s\in \mathcal{S}$.
\FOR {each time step $t=1,2,\ldots,T$}
\STATE Observe state $s^t$ according to the wireless environment interaction;
\STATE Choose action $a^t$ according to the $\varepsilon$-greedy policy;
\STATE Perform action $a^t$, observe the state $s^{t+1}$ and obtain the reward $r^t$ feedback from the environment;
\STATE Store the experience $e(s^t,a^t,r^t,s^{t+1})$ in replay memory $\mathcal{D}$;
\STATE Sample a random minibatch of $|\mathbf{e}|$ experiences from replay memory $\mathcal{D}$;
\STATE Calculate the target Q-value by using \eqref{eq:yt};
\STATE Train the evaluated network $\eta$ by performing a gradient descent step \eqref{eq:decent};
\STATE Update the weight parameter $\eta$ of the evaluated network;
\STATE Update the state $s^t\leftarrow s^{t+1}$;
\ENDFOR
\STATE Update the weight parameter $\eta^-$ of the target network every $S$ steps;
\ENDFOR
\STATE \textbf{\underline{DEPLOYMENT STAGE}:}
\FOR {each time step $t=1,2,\ldots$}
\STATE Observe the state $o^t$ according to the real-world wireless environment interaction;
\STATE Perform the optimized ST strategy with respect to the phase-shift of RIS.
\ENDFOR
\end{algorithmic}
\end{algorithm}

\subsection{System Complexity}
\subsubsection{Computational Complexity}
The computational complexity of the proposed algorithm is evaluated by the floating point operations per second (FLOPS), which counts addition, subtraction, multiplication, division, exponentiation, etc., as a single FLOP. Specifically, the total computational complexity includes training complexity and deployment complexity, which are analyzed in the following.
 \begin{enumerate}
   \item Training complexity: First, we derive the computations of activation layers. The computations are ``1" for Relu layer, ``4" for sigmoid layer and ``6" for tanh layer. We assume that the total nodes for state normalization layer, Relu layer, sigmoid layer and tanh layer are $|\mathcal{S}|$, $\tilde{n}_r$, $\tilde{n}_s$ and $\tilde{n}_t$, respectively. Thus, the training complexity for node computation can be calculated by $\mathcal{O}(|\mathcal{S}|+\tilde{n}_r+4\tilde{n}_s+6\tilde{n}_t)$. Furthermore, we assume that both evaluated network and target network are $M$ fully connected layers in total and the $m$th layer has $\tilde{n}_m$ nodes. The training complexity for one forward propagation and two backward propagation can be calculated  by $\mathcal{O}(\sum_{m=0}^{M-1}3\tilde{n}_m\tilde{n}_{m+1})$. Therefore, the training complexity of the evaluated network for $T$ time steps is $\mathcal{O}\big(T\cdot|\mathbf{e}|\cdot((|\mathcal{S}|+\tilde{n}_r+4\tilde{n}_s+6\tilde{n}_t)+\sum_{m=0}^{M-1}3\tilde{n}_m\tilde{n}_{m+1})\big)$ with samples $|\mathbf{e}|$ form replay buffer. On the other hand, the training complexity of the target network for $T$ time steps is $1/S$ of the former. Therefore, for $\mathcal{E}$ episodes, the training complexity of the proposed DRQN based dynamic ST strategy can be approximated as $\mathcal{O}\big(\mathcal{E}\cdot T\cdot|\mathbf{e}|\cdot(\sum_{m=0}^{M-1}\tilde{n}_m\tilde{n}_{m+1}+|\mathcal{S}|)\big)$.
   \item Deployment complexity: For the deployment stage, there is no target network and replay buffer used, thus only the computation complexity of the state normalization and one forward propagation are considered. In this case, for either Alice or Bob, the total deployment complexity is approximated as $\mathcal{O}(\sum_{m=0}^{M-1}\tilde{n}_m\tilde{n}_{m+1})+\mathcal{O}(|\mathcal{S}|)$.
 \end{enumerate}
\begin{remark}
The computational complexity of the proposed DRQN based dynamic ST strategy mostly depends on the weight parameter updating for forward and backward propagation. In addition, for a fixed neural network structure, the computational complexity is increased linearly with the number of episodes $\mathcal{E}$, time steps $T$, the size of the minibatch sample $|\mathbf{e}|$ and state space $|\mathcal{S}|$, which indicates that the running time is polynomial.
\end{remark}
\subsubsection{Hardware Complexity} The hardware complexity of the proposed algorithm is discussed from the power consumption and the hardware resource aspects. To achieve an excellent ICAS performance, significant power consumption and hardware resources are required for the offline training with a state set as large as possible, including the communication consumption, RIS configuration, network forward propagation and backword propagation, and step iteration among others. Nevertheless, in the online deployment, this hardware complexity can be substantially reduced, where the power source and the hardware resources are mainly used to support the communication consumption, network forward propagation
and RIS configuration. Thus, the hardware complexity is acceptable when the trained neural network is deployed online. However, integrating the proposed algorithm into the hardware platform still faces some challenges in practice, which is due to the limitations of the current hardware technology. For example, the computing ability of the artificial intelligence (AI) chips for exploring the real-world unknown boundary state set is not strong enough, while the real-time processing capabilities of FPGAs for configuring RISs to construct a fast time-varying channel environment are also weak.

\section{Simulation Results}\label{sec:5}
In this section, the performance of the proposed algorithm is evaluated via simulations. We first provide the details of the simulation setup, and then analyze the simulation results.
\subsection{Simulation Setup}
We utilize the compute unified device architecture (CUDA) in PyTorch as the computational software platform and the NVIDIA GTX 1660 Ti as the computational hardware platform. The high-quality computational platforms can accelerate the network training, and then contribute to the smoother simulations. For the structure of the neural network, both the target network and the evaluated network are designed with three layers. The input and output layers are linear layers with the dimensions corresponding to the state space and the action space, respectively. Moreover, an intermediate LSTM layer is utilized for memory retention. The ReLU activation function is applied in each layer of the network. Specifically, for the end-to-end data training, the input is first subjected to a linear transformation through the Linear1 layer, followed by activation through ReLU function. Then, the result is fed into the LSTM layer, generating the output, along with the new LSTM hidden state and the cell state. Finally, the LSTM output is fed into the output layer Linear2, yielding the output of the network, which represents the estimated Q-value. The main parameter settings are shown in Table \ref{tab:2}.
\begin{table}[htbp]
\centering
\caption{The main parameter settings of the developed network}\label{tab:2}
\renewcommand\arraystretch{1.2}
\setlength{\tabcolsep}{6mm}{
\begin{tabular}{l|l}
\hline
\textbf{Parameters} & \textbf{Value} \\
\hline
Size of hidden layer & 64 \\

Learning rate & 0.1 \\

Batch size & 32 \\

Episodes & 1500 \\

Max episode step & 20\\

Starting epsilon & 1.0 \\

End epsilon & 0.001 \\

Epsilon decay rate & 0.995 \\
Max episode step & 20\\

Random update & True \\

Optimizer & SGD \\

Neural network layer & 2Linear, 1LSTM \\

Activation function & ReLU \\
\hline
\end{tabular}
}
\end{table}

For the channel environments, we set three types of arrays which are the diagonal phase-shift matrix, the channel gain vector between RIS to users and the direct link channel gain of user. The dimensions of them are $N\times N$, $N\times1$, and $1\times1$, respectively. The diagonal matrix of the RIS phase-shift is generated from the range $(0, 2\pi]$ with the finite resolution of $b$ bits. The other matrices for the channel gains are generated using the random variables with complex Gaussian distribution. The bandwidth of the transmit signal is set to be $B=1$ Hz and the time for each channel estimation round is set to be $T_s=1$ s. Based on this, the real and imaginary parts of the matrices are stored in separate arrays, which establishes the state space. The dimension of the state space is related to the number of RIS reflecting elements and the dimension of each channel.
The reward is calculated by the expressions of the data transmission rate and the key generation rate.
It is worth noting that the calculated reward may exhibit some deviation due to the use of different random generation data for the state space, however, this does not impact the performance evaluation of the proposed DRQN based dynamic ST strategy. Furthermore, to mitigate the effect of noise in the experimental results, we utilized a moving average function to smooth the reward curve. Furthermore, the max episode step is defined as the maximum number of steps allowed per episode.

\subsection{Performance Analysis}
\subsubsection{Performance of Network Convergence}
\begin{figure}[!ht]
 \centering
  \includegraphics[width=8cm]{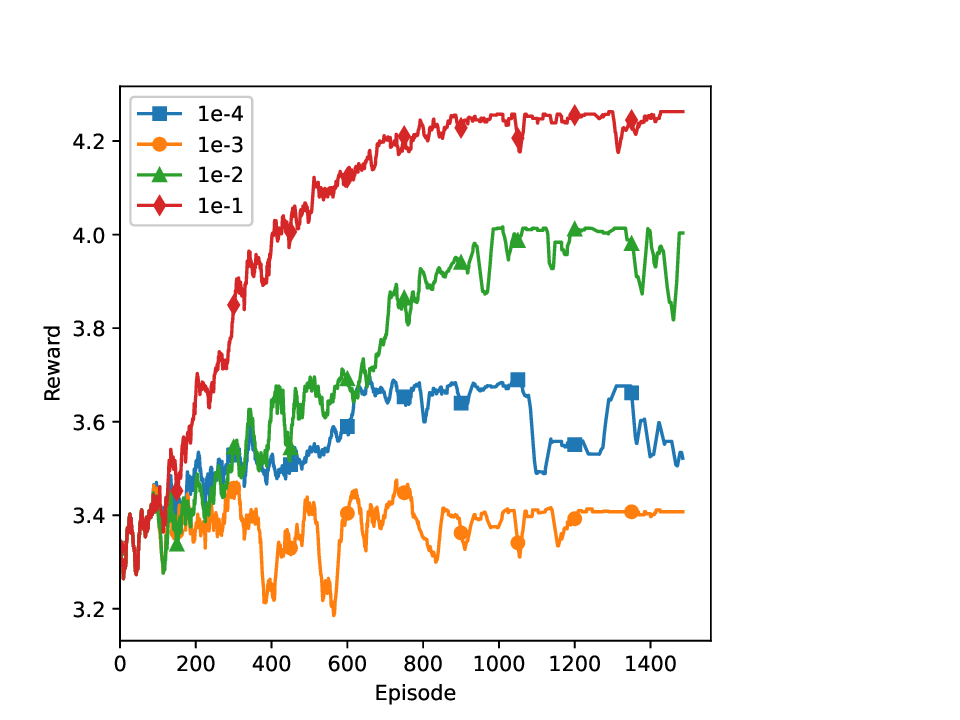}
  \caption{The convergence of the proposed DRQN based dynamic ST strategy for different learning rates.}
  \label{Fig:F1a}
\end{figure}

For the performance analysis of the network, the action space consists of a randomly generated set of actions with the dimensions $8\times8$, which corresponds to a continuous finite RIS phase-shift. This implies that we have 8 actions to choose from, where each action comprises the phase-shift of 8 RIS reflecting elements.
In Fig. \ref{Fig:F1a}, we study the convergence of the proposed  DRQN based dynamic ST strategy for different learning rates. By observing all reward curves,  we can see that the reward curves fluctuate largely at the beginning of the 600 episodes, and then, the fluctuations of the reward curves tend to flatten out. When the episodes reach to about 800, each reward curve gradually converges to a fixed point but still with a slight fluctuation. With the increase of the episodes, the trend of the convergence is more obvious. For example, when the number of episodes is 1500, the reward curve with learning rate $0.1$ converges to 4.27 bits/s. On the other hand, by observing each reward curve, it can be found that the convergence value of the reward curve is not proportional to the decrease of the learning rate. If the learning rate is not suitable for the state space, the fluctuation of the Q-function is very large and the evaluated network cannot accurately estimate the reward feedback form the Q-function. As a result, the evaluated network can fall into a local optimal strategy at the end of the episode. Therefore, it is very important to select a proper learning rate for network training.

\begin{figure}[!ht]
 \centering
  \includegraphics[width=8cm]{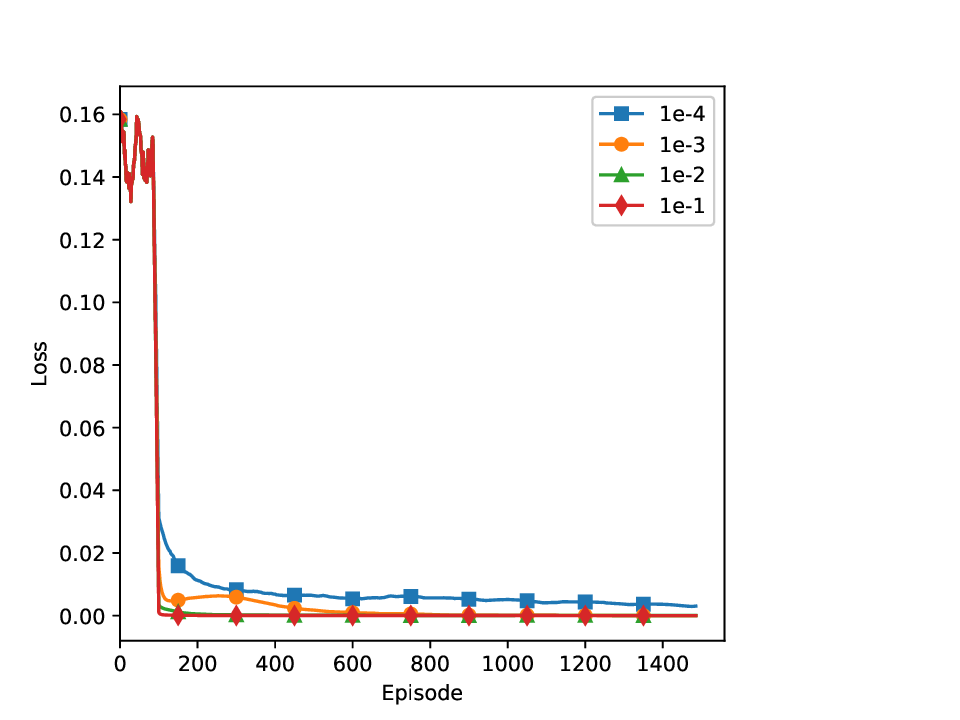}
  \caption{The variation of the loss function with respect to the number of episodes for different learning rates.}
  \label{Fig:F1b}
\end{figure}

Figure \ref{Fig:F1b} illustrates that the value of the loss function fluctuates in the first 100 episodes and then decreases rapidly in the following episodes. The reason is that at the beginning of the iterations, the evaluated network is mainly exploring and exploiting the state space, which causes the fluctuation of the loss function within a specific range. However, after this period, the exploiting process of the evaluated network is dominant, and the value of the loss function decreases significantly with the number of the episodes. Specifically, when the learning rate reduces, the final value of the loss function increases. This shows that a small learning rate may cause overfitting of the neural network and leading to a local convergence value. Furthermore, with the increase of the episodes, we can observe that the values of the loss function for learning rates $0.1$, $0.01$ and $0.001$ are almost overlapped. However, small differences in the value of the loss function can cause large differences in convergence performance. This insight also demonstrates the importance of selecting a suitable learning rate for network training.

\subsubsection{Performance of Proposed Strategy}
\begin{figure}[!ht]
 \centering
  \includegraphics[width=8cm]{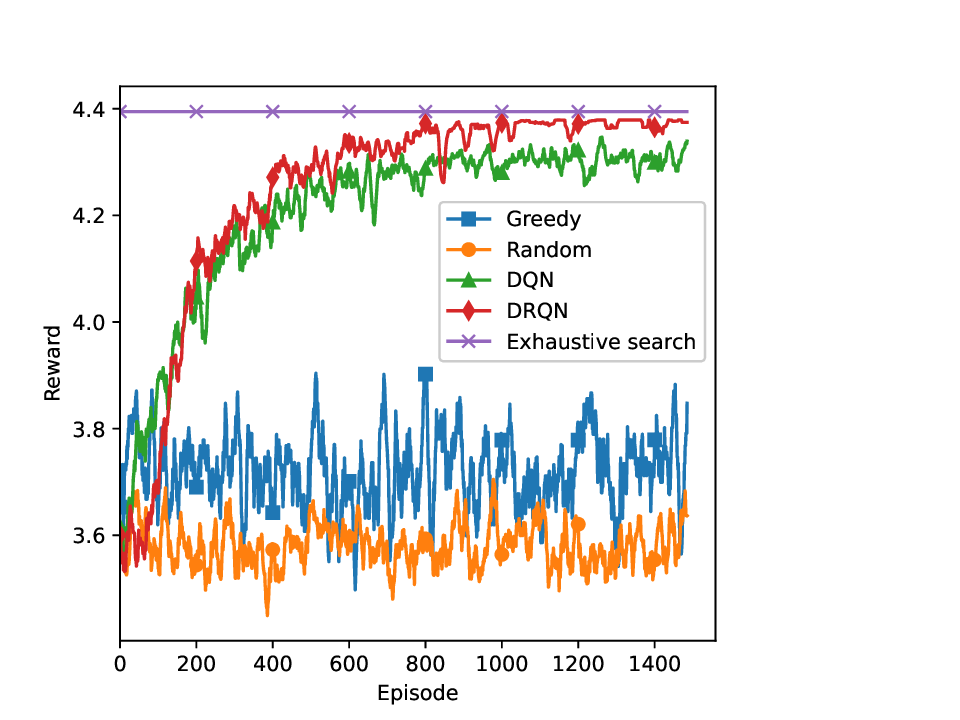}
  \caption{The reward of the proposed DRQN based
dynamic ST strategy vs. four benchmarks.}
  \label{Fig:F2}
\end{figure}

Based on the above network parameter settings, we now compare the performance of the proposed DRQN based dynamic ST strategy with four benchmarks, which are the DQN based strategy, the greedy strategy, the random strategy and the exhaustive search strategy. From Fig. \ref{Fig:F2}, it is straightforward to see that the exhaustive search strategy is optimal since it sweeps the whole action space in each episode. We can observe that the performance of the proposed DRQN based dynamic ST strategy is almost consistent with the optimal exhaustive search strategy. Notably,
although the exhaustive search strategy can achieve a desirable performance, the computational overhead is very large. For the proposed DRQN based dynamic ST strategy, when the neural network is well trained, we can deploy the evaluated network to perform the dynamic ST strategy, which has a low overhead. Thus, our proposed strategy has a strong advantage in optimizing the dynamic ST strategy. Furthermore, it can be found that the performance of the proposed DRQN based dynamic ST strategy is better than of both the greedy strategy and the random strategy. Since the random strategy lacks the interaction with the wireless environment, the performance of the random strategy is the worst among the benchmarks.
\begin{figure}[!ht]
 \centering
  \includegraphics[width=8cm]{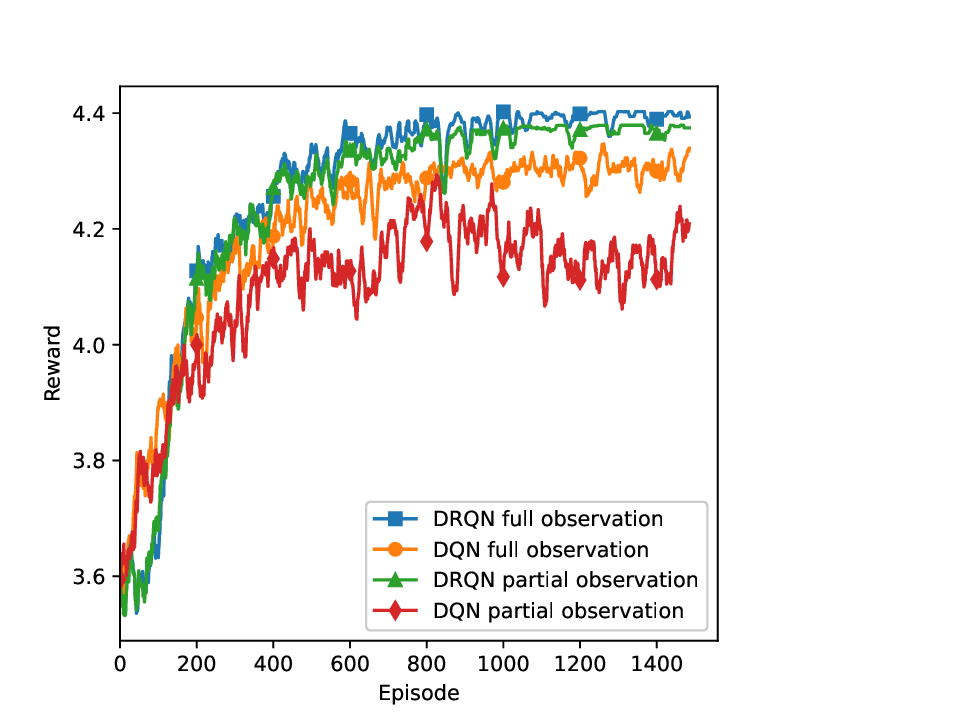}
  \caption{The reward of various deep reinforcement learning based dynamic ST strategies with full observation and partial observation information.}
  \label{Fig:F3}
\end{figure}
We now analyze the robustness of the  proposed DRQN based dynamic ST strategy in Fig. \ref{Fig:F3}. Specifically, we use the full observation information and the partial observation information to test the reward of the trained DRQN based dynamic ST strategy and DQN based strategy. From the figure, it can be found that the rewards of the proposed strategy for full observation information and partial observation information are about 4.4 bit/s and 4.38 bit/s, respectively. Moreover, the rewards of the DQN based strategy for full observation information and partial observation information are about 4.31 bit/s and 4.15 bit/s, respectively. The result shows that although the performance of both ST strategies deteriorates, our proposed DRQN based dynamic ST strategy also has a better performance than the DQN based strategy with only partial observation information, i.e., the performance loss are 0.02 bit/s and 0.16 bit/s, respectively. The essential reason is that the LSTM layer of the DRQN can learn the history relations of the states in time series by the recurrent control. When the state cannot be full observed, the trained DRQN still can infer more dynamic relations of the state form the partial observation information than the trained DQN, thereby maintaining a better performance.

\begin{figure}[!ht]
 \centering
  \includegraphics[width=8cm]{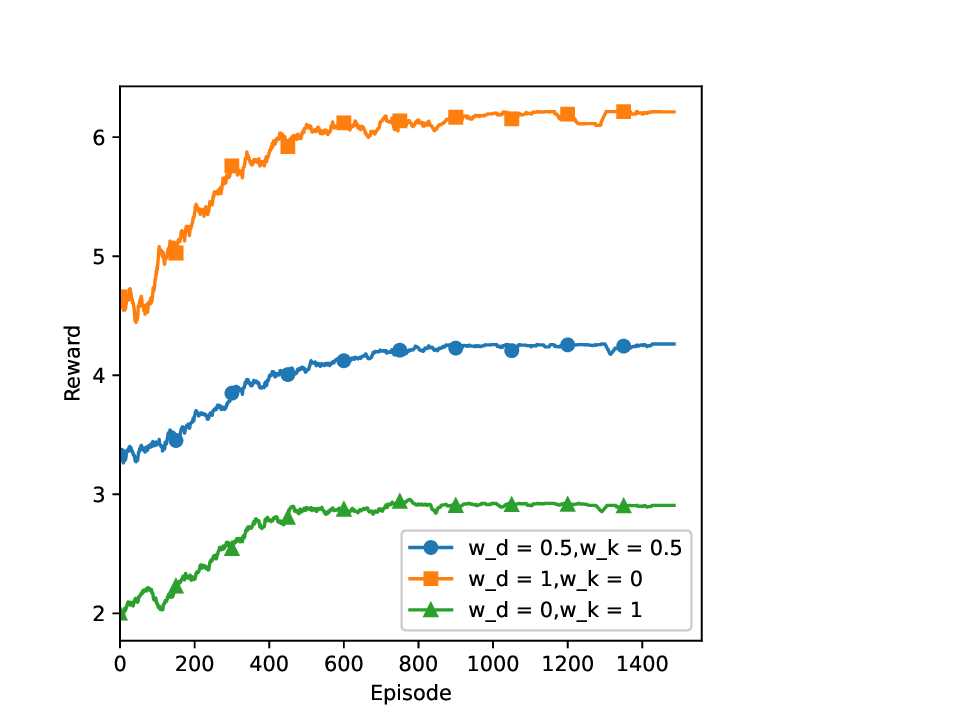}
  \caption{The reward of the proposed DRQN based dynamic ST strategy with different weight factors.}
  \label{Fig:F4a}
\end{figure}
Next, the rules of weight factors for the proposed DRQN based dynamic ST strategy are analyzed. Figure \ref{Fig:F4a} shows the reward of the proposed DQN based strategy with different weight factor trade-offs. For the first case, when the ICAS design tends to be promote the communication mode, i.e., $w_d=1,w_k=0$, we can find that the reward is about 6.19 bit/s at the convergence point. In this case, the phase-shift is optimized to maximize the data transmission rate, thus, the key generation rate is low. For the second case, when the weight factor trade-off is $w_d=0,w_k=1$, the design objective is to maximize the key generation rate without considering optimizing the data transmission rate. We observe that the reward is about 2.89 bit/s. For the third case, when both the data transmission rate and the key generation rate are considered, such as the weight factor trade-off is $w_d=0.5,w_k=0.5$, the phase-shift is optimized to maximize the data transmission rate and the key generation rate, simultaneously. We infer that the reward of the third case is between the first case and the second case which represent the upper bound and the lower bound of the proposed DRQN based dynamic ST strategy, respectively. Therefore, there is an interesting performance tradeoff between the data transmission rate and the key generation rate by allocating the weight factor. The trade-off principle of the weight factor depends on whether the design purpose is to conduct the data transmission task or the secret key generation task.

\begin{figure}[!ht]
 \centering
  \includegraphics[width=8cm]{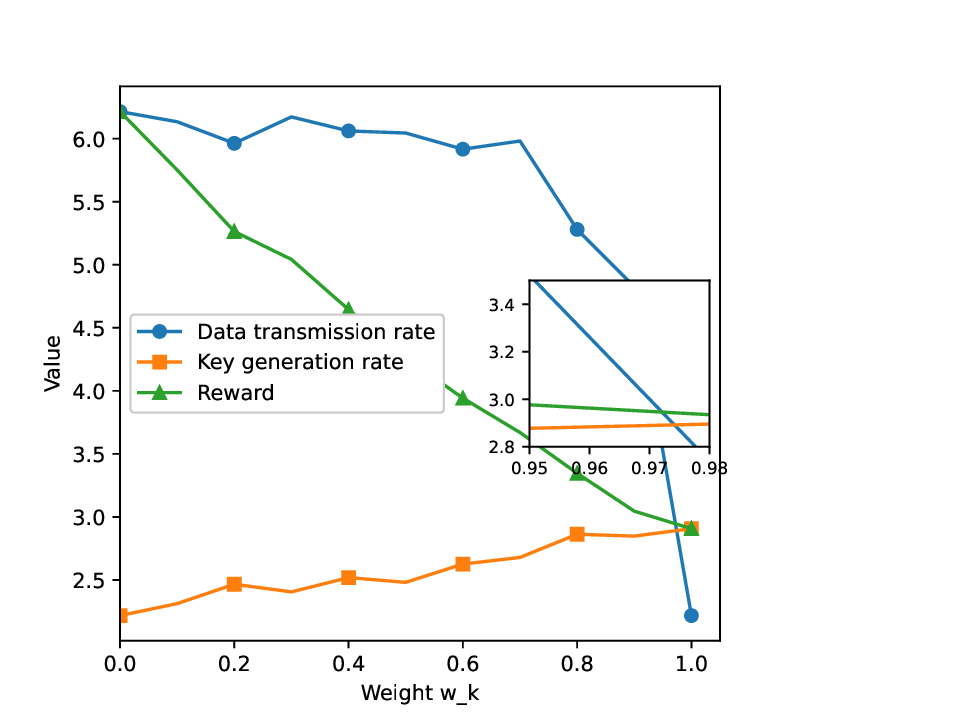}
  \caption{The variations of the data transmission rate and the key generation rate with respect to different weight factors.}
  \label{Fig:F4b}
\end{figure}

The result in Fig. \ref{Fig:F4b} shows the influence of the weight factor $w_k$ on the optimal data transmission rate and the key generation rate, respectively. We observe that different weight factor trade-offs can lead to different design purposes for simultaneous transmission and secret key generation. For example, when the weight factor is $w_k=0$, the design purpose is to maximize the data transmission rate. Similarly, when the weight factor is $w_k=1$, the design purpose is to maximize the key generation rate. The rewards of these cases are consistent with the result in Fig. \ref{Fig:F4a}. Changing the weight factor $w_k$ from 0 to 1, it can be observed that the reward is almost a monotonically decreasing function with respect to weight $w_k$. However, we can further observe that the variation trend of the key generation rate is not exactly proportional to that of the data transmission rate. Particularly, when the weight factor is $w_k\in(0.6,0.7)$, both the data transmission rate and the key generation rate are improved, simultaneously. This observation implies that the ICAS design of simultaneously transmitting and generating secret key produces the effect of mutualism. When the weight factor $w_k$ changes from 0 to 0.9, we find that the transmission rate is decreased sharply from $w_k=0.7$, yet it remains always higher than the key generation rate. Thus, the key generation rate cannot satisfy the requirements of the data transmission rate for ``one time pad" communication. As a consequence, selecting an appropriate weight factor trade-off is an important task to achieve a ``one time pad" communication. Interestingly, when the weight factor $w_k$ is between 0.9 to 1, it can be found that the data transmission rate can be equal to the key generation rate, which indicates that the ``one time pad" communication is possible via an ICAS design. For instance, in our simulation settings, we notice that $w_k=0.973$ is a suitable weight factor for designing simultaneous transmission and generation of secret key to realize the ``one time pad" communication. Although this ``one time pad" communication compromises a part of the communication performance, it is very necessary in some scenarios with strong security requirements, i.e., military communications.
\subsubsection{Performance of Different RIS Parameters}
\begin{figure}[!ht]
 \centering
  \includegraphics[width=7.8cm]{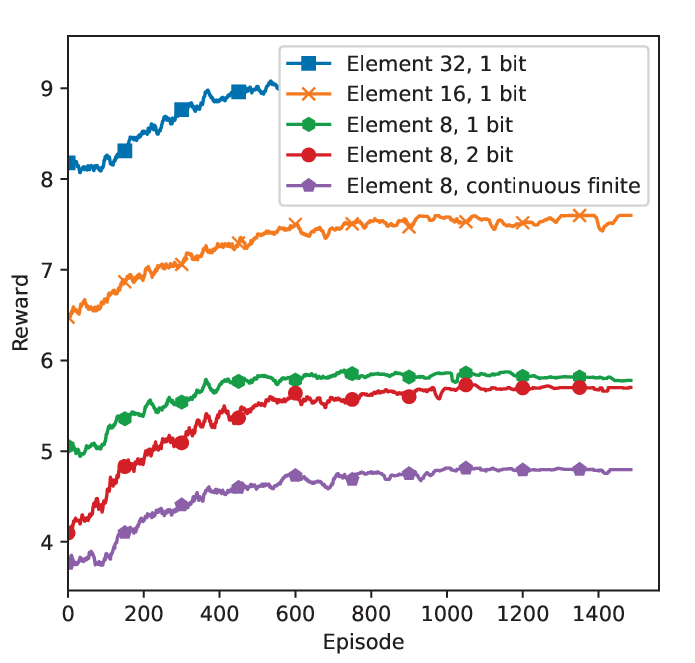}
  \caption{The system reward with different RIS parameter settings.}
  \label{Fig:F11}
\end{figure}

As the RIS is the pivotal hardware platform for the realization of simultaneous transmission and generation of secret keys, we analyze the different RIS parameters effect on the proposed DRQN based dynamic ST strategy, including the number of RIS refection elements and the finite resolution of the RIS phase-shift, respectively. To ensure the fairness of our performance analysis, we use the same parameter settings of the network, except from the state space, action space and the finite resolution which are populated due to the RIS parameters settings. It can be observed from Fig. \ref{Fig:F11} that with the increase of the RIS refection elements, the reward increases significantly. A larger RIS can improve the system reward, but also entails heavier energy consumption and cost overhead. For the finite resolution of the RIS phase-shift, we notice that the convergence rewards of 1 bit resolution and 2 bit resolution are almost consistent. This interesting result is due to that both the 1 bit and 2 bit resolutions are lower, and thus have the same effect on the system reward improvement. Theoretically, the continuous RIS phase-shift can significantly increase the reward due to the largest phase-shift degrees of freedom. However, as shown in the figure, since the DRQN framework cannot deal with the continuous action, the reward of the continuous finite RIS phase-shift can be lower than that of the 1 bit and 2 bit resolutions, which is attributed to the sampling data of the action space.

\section{Conclusion}\label{sec:6}
We have studied the potential of simultaneous transmission and
generation of secret key for RIS-assisted ICAS system in the presence of a smart attacker, where the phase-shift of the RIS has been optimized. Specifically, we formulated the problem as a ST game for the dynamic wireless environments, and proved the existence of the NE point. Next, considering the snapshot of the ST game, we derived the NE point of the static game. On the other hand, considering the whole evolutionary process of the
ST game, we proposed a DRQN based dynamic ST strategy to learn the optimal phase-shift of RIS. Then, we provided the details
of the algorithm and analyzed the system
complexity. The simulation results demonstrated that the proposed DRQN based dynamic ST strategy can approach the optimal exhaustive search strategy. In particular, it has been proved that the proposed DRQN based dynamic ST strategy has a better performance than the benchmarks even with partial observation information. By allocating the suitable weight factor to balance the data transmission and the secret key generation, we have shown that the ``one time pad" communication becomes possible.
\appendices
\section{Proof of Proposition 1}\label{sec:AP1}
The channel cross-correlation among transmitter $\tilde{t}$ and two receivers $r1,r2$ can be expressed as
\begin{align}
R_{r1,r2}=\mathbb{E}\{(\mathbf{h}^T_{Rr1}\mathbf{\Phi}\mathbf{h}_{\tilde{t}R}
+h_{\tilde{t}r1})(\mathbf{h}^T_{Rr2}\mathbf{\Phi}\mathbf{h}_{\tilde{t}R}
+h_{\tilde{t}r2})^H\}~~\nonumber\\
=\mathbb{E}\{(\mathbf{h}^T_{Rr1}\mathbf{\Phi}\mathbf{h}_{\tilde{t}R})(\mathbf{h}^T_{Rr2}\mathbf{\Phi}\mathbf{h}_{\tilde{t}R})^H\}+\mathbb{E}\{h_{\tilde{t}r1}h_{\tilde{t}r2}^*\}.
\end{align}
Let the cascaded channels be
\begin{align}
\mathbf{h}^T_{Rr1}\mathbf{\Phi}\mathbf{h}_{\tilde{t}R}=\mathbf{v}^T_\theta\mathbf{\Lambda}_{Rr1}\mathbf{h}_{\tilde{t}R}\nonumber\\
\mathbf{h}^T_{Rr2}\mathbf{\Phi}\mathbf{h}_{\tilde{t}R}=\mathbf{v}^T_\theta\mathbf{\Lambda}_{Rr2}\mathbf{h}_{\tilde{t}R},
\end{align}
where $\mathbf{\Lambda}_{Rr1}=\text{diag}(\mathbf{h}^T_{Rr1})$, and $\mathbf{\Lambda}_{Rr2}=\text{diag}(\mathbf{h}^T_{Rr2})$. Then, the cross-correlation of the cascaded channel can be rewritten as
\begin{align}
\mathbb{E}\{(\mathbf{v}^T_\theta\mathbf{\Lambda}_{Rr1}\mathbf{h}_{\tilde{t}R})(\mathbf{v}^T_\theta\mathbf{\Lambda}_{Rr2}\mathbf{h}_{\tilde{t}R})^H\}~~~~~~\nonumber\\
=\mathbb{E}\{(\mathbf{v}^H_\theta\mathbf{\Lambda}^*_{Rr2}\mathbf{h}^*_{\tilde{t}R})(\mathbf{v}^T_\theta\mathbf{\Lambda}_{Rr1}\mathbf{h}_{\tilde{t}R})^T\}~~~~~\nonumber\\
=\mathbb{E}\{\mathbf{v}^H_\theta\mathbf{\Lambda}^*_{Rr2}\mathbf{h}^*_{\tilde{t}R}\mathbf{h}^T_{\tilde{t}R}\mathbf{\Lambda}^T_{Rr1}\mathbf{v}_\theta\}.~~~~~~~~~~~
\end{align}
Thus, the cross-correlation between the transmitter $\tilde{t}$ and two receivers $r1,r2$ is
\begin{align}
R_{r1,r2}=\mathbf{v}^H_\theta\mathbf{R}_{r1,r2}\mathbf{v}_\theta+R^D_{r1,r2},
\end{align}
where the symbols $\mathbf{R}_{r1,r2}=\mathbb{E}\{\mathbf{\Lambda}_{Rr2}^*\mathbf{h}^*_{\tilde{t}R}\mathbf{h}^T_{\tilde{t}R}\mathbf{\Lambda}^T_{Rr1}\}$ and $R^D_{r1,r2}=\mathbb{E}\{h_{\tilde{t}r1}h_{\tilde{t}r2}^*\}$. When $r1=r2$, we can calculate the channel variance of between the transmitter and receiver $r1$, $r2$ as
\begin{align}
R_{r1}=\mathbf{v}^H_\theta\mathbf{R}_{r1}\mathbf{v}_\theta+R^D_{r1},\nonumber\\
R_{r2}=\mathbf{v}^H_\theta\mathbf{R}_{r2}\mathbf{v}_\theta+R^D_{r2},
\end{align}
where the symbols $\mathbf{R}_{r1}=\mathbb{E}\{\mathbf{\Lambda}_{Rr1}^*\mathbf{h}^*_{\tilde{t}R}\mathbf{h}^T_{\tilde{t}R}\mathbf{\Lambda}^T_{Rr1}\}$, $\mathbf{R}_{r2}=\mathbb{E}\{\mathbf{\Lambda}_{Rr2}^*\mathbf{h}^*_{\tilde{t}R}\mathbf{h}^T_{\tilde{t}R}\mathbf{\Lambda}^T_{Rr2}\}$, while $R^D_{r1}=\mathbb{E}\{h_{\tilde{t}r1}h_{\tilde{t}r1}^*\}$, $R^D_{r2}=\mathbb{E}\{h_{\tilde{t}r2}h_{\tilde{t}r2}^*\}$, respectively.
\section{Proof of Proposition 2}\label{sec:AP2}
\begin{definition}
A set $\mathcal{P}\subset \mathbb{R}^2$ is a convex set if for every $P',P''\in\mathcal{P}$ and $\rho\in[0,1]$, it satisfies
$\rho P'+(1-\rho)P''\in\mathcal{P}$.
\end{definition}
\begin{definition}
A set $\mathcal{P}\subset \mathbb{R}^2$ is called compact if the set is closed and bounded \cite{jiang2009tutorial}.
\end{definition}
\begin{lemma}
\label{lem:1}
Let a function $f:\mathcal{P}\rightarrow\mathcal{P}$ be continuous, then there exists some $P'\in\mathcal{P}$, satisfying $f(P)=P'$, that is, the function $f$ has one fixed point.
\end{lemma}
For $i\in\{A,E\}$ and a given mixed strategy profile $P\in \mathcal{P}$, we can define $\varphi_{i,a_i}(P)=\max\{0,\bar{u}_i(a_i,P_{-i})-\bar{u}_i(P)\}$, where $P_{-i}$ is the strategy profile of all participants except $i$. According to Lemma \ref{lem:1}, we formulate the function $f:\mathcal{P}\rightarrow\mathcal{P}$ as $f(P)=P'$, where
\begin{align}
P_i'(a_i)&=\frac{P_i(a_i)+\varphi_{i,a_i}(P)}{\sum_{\tilde{a}_i\in \mathcal{A}_i}(P_i(\tilde{a}_i)+\varphi_{i,\tilde{a}_i}(P))}\\\nonumber
&=\frac{P_i(a_i)+\varphi_{i,a_i}(P)}{1+\sum_{\tilde{a}_i\in \mathcal{A}_i}\varphi_{i,\tilde{a}_i}(P)}.
\end{align}
The above equation suggests that $0\leq P_i'(a_i)\leq1,\forall i\in\{A,E\}$ and $f$ is a self mapping. This means that the actions that have better response to $P$ can increase its probability distributions. In the formulated ST game,
since $i\in\{A,E\}$, the set of the mixed strategy $\mathcal{P}\in \mathbb{R}^2$ is in a closed and bounded rectangular region, which is convex and compact. On the other hand, the function $f$ is continuous since each $\varphi_{i,a_i}$ is continuous. Thus, the function $f$ has one fixed point.

Next, we prove that the fixed points of $f$ are the Nash equilibria. Obviously, when the mixed strategy profile $P$ is a Nash equilibrium, we have $\varphi_{i,a_i}(P)=0, \forall i\in\{A,E\}$, which shows that $P$ is one fixed point of $f$. On the contrary, assuming $P$ is an arbitrary fixed point of $f$, there must exist at least one action $a'_i$ in the support of $P$, satisfying $\bar{u}_i(a'_i,P_{-i})\leq \bar{u}_i(P)$. From the definition of $\varphi$, we have $\varphi_{i,a'_i}(P)=0$. Moreover, as per the property of the fixed point, we have $P_i'(a_i')=P_i(a_i')$, which implies that the denominator must be 1 and $\varphi_{i,a_i}(P)$ must be 0,$\forall i\in\{A,E\}, \tilde{a}_i\in \mathcal{A}_i$, that is $\sum_{\tilde{a}_i\in \mathcal{A}_i}\varphi_{i,\tilde{a}_i}(P)=0$. These conditions are confirmed only when no participant can improve the expected utility by moving to a pure strategy. Then, the proof is completed.
\section{Proof of Proposition 3}\label{sec:AP3}
If $(\mathbf{v}_\theta^\star,0)$ is a NE point for the static ST game, we have
\begin{eqnarray}
u_E(\mathbf{v}_\theta^\star,0)-u_E(\mathbf{v}_\theta^\star,1)=0-(\mathcal{R}_{k,E}|_{\mathbf{v}_\theta=\mathbf{v}_\theta^\star}-C_E)\geq0.
\end{eqnarray}
Thus, this inequality condition supports $u_E(\mathbf{v}_\theta^\star,0)\geq u_E(\mathbf{v}_\theta^\star,\kappa),
~\forall \kappa\in\{0,1\}$.
After algebraic manipulations, the data transmission rate in \eqref{eq:regodic} can be rewritten as
\begin{align}
\label{eq:zRd}
&~~~\mathcal{R}_d\nonumber\\
&=B\log_2(1+\mathbb{E}\{(\mathbf{v} ^T_\theta\mathbf{\Lambda}_{RA}\mathbf{h}_{BR})(\mathbf{v}^T_\theta\mathbf{\Lambda}_{RA}\mathbf{h}_{BR})^H+h_{BA}h_{BA}^*\})\nonumber\\
&=B\log_2(1+\mathbf{v}^H_\theta \mathbf{R}_A\mathbf{v}_\theta+R_A^D).
\end{align}
Next, let $z=\mathbf{v}^H_\theta \mathbf{R}_A\mathbf{v}_\theta+R_A^D$. Substituting this into \eqref{eq:Krate1} and \eqref{eq:zRd}, the function of optimization objective $u_A(\mathbf{v}_\theta,0)$ can be denoted as
\begin{eqnarray}
f(z)=w_dB\log_2(1+z)+\frac{w_k}{T_s}\log_2\bigg(\frac{(1+z)^2}{1+2z}\bigg).
\end{eqnarray}
Then, the first derivative of $f(z)$ with respect to $z$ is derived as
\begin{eqnarray}
\frac{\partial f(z)}{\partial z}=\frac{1}{\ln2}\frac{2w_dBz+2\frac{w_k}{T_s}z+w_dB}{(z+1)(2z+1)}.
\end{eqnarray}
Obviously, for $z\geq0$, we can find that $\frac{\partial f(z)}{\partial z}\geq0$ always holds, which indicates that the function $f(z)$ is monotonically increasing with $z\geq0$. On the other hand, $\forall \mathbf{v}_\theta$, we have $\mathbf{v}^H_\theta \mathbf{R}_A\mathbf{v}_\theta\geq0$, which indicates that $\mathbf{R}_A\succ \mathbf{0}$ is a positive semi-definite Hermitian matrix. For every non-zero $\mathbf{v}$ in a Rayleigh quotient, we have the following inequality
\begin{eqnarray}
\label{eq:Rayleigh}
\lambda_{\min}(\mathbf{R}_A)\leq\frac{\mathbf{v}^H \mathbf{R}_A\mathbf{v}}{\mathbf{v}^H\mathbf{v}}\leq \lambda_{\max}(\mathbf{R}_A),
\end{eqnarray}
where $\lambda_{\min}(\mathbf{R}_A)$ and $\lambda_{\max}(\mathbf{R}_A)$ are the minimum and maximum eigenvalues of matrix $\mathbf{R}_A$, respectively. For a given matrix, the Rayleigh quotient reaches its maximum value $ \lambda_{\max}(\mathbf{R}_A)$ when $\mathbf{v}$ corresponds to the maximum eigenvector, i.e., $\mathbf{v}=\mathbf{v}_{\max}$ with $\mathbf{v}_{\max}^H\mathbf{v}_{\max}=1$. Then, by setting $\mathbf{v}^{\star}_\theta=\sqrt{N}\mathbf{v}_{\max}$,
it is straightforward to show that the maximum value of $z$ can be obtained when $\mathbf{v}^{\star H}_\theta\mathbf{v}^\star_\theta=N\mathbf{v}_{\max}^H\mathbf{v}_{\max}$. By substituting $\mathbf{v}^{\star H}_\theta\mathbf{v}^\star_\theta$ into \eqref{eq:Rayleigh}, we have
\begin{eqnarray}
\label{eq:lamdamax}
\mathbf{v}^{\star H}_\theta \mathbf{R}_A\mathbf{v}^\star_\theta=N\lambda_{\max}(\mathbf{R}_A).
\end{eqnarray}
As a consequence, when \eqref{eq:lamdamax} holds and $C_E\geq \mathcal{R}_{k,E}|_{\mathbf{v}_\theta=\mathbf{v}_\theta^\star}$, we have
\begin{eqnarray}
&&u_A(\mathbf{v}_\theta^\star,0)\geq u_A(\mathbf{v}_\theta,0),~\forall0\leq\theta_n\leq2\pi\nonumber\\
&&u_E(\mathbf{v}_\theta^\star,0)\geq u_E(\mathbf{v}_\theta^\star,\kappa),
~\forall \kappa\in\{0,1\}.
\end{eqnarray}
The proof is completed.
\ifCLASSOPTIONcaptionsoff
  \newpage
\fi



\bibliographystyle{IEEEtran}
\bibliography{IEEEabrv,sigproc} 
\end{document}